\documentclass[reqno,12pt]{amsart}
\usepackage{fullpage}

\usepackage{amssymb,amsmath,amsfonts,latexsym}
\usepackage{amsthm}

\usepackage{graphicx}
\usepackage{enumerate}
\usepackage{url}

\numberwithin{equation}{section}
\def\Rnum{\mathbb{R}}
\def\Mink#1{\Rnum^{#1,1}}

\def\pos{{\vec \gamma}}
\def\T{{\mathbf T}}
\def\N{{\mathbf N}}
\def\B{{\mathbf B}}

\def\e{{\mathbf e}}
\def\E{{\mathbf E}}
\def\U{{\mathbf U}}
\def\W{{\mathbf W}}
\def\H{{\mathbf H}}
\def\A{{\mathbf A}}

\def\S{{\mathbf S}}

\def\lieder#1{{\mathfrak L}_{#1}}
\def\X{{\mathrm X}}

\def\spn{\text{span}}
\def\t{{\rm t}}

\def\Rop{{\mathcal R}}
\def\Qop{{\mathcal Q}}
\def\Pop{{\mathcal P}}
\def\Kop{{\mathcal K}}

\def\Dop{{\mathcal D}}
\def\Eop{{\mathcal E}}
\def\Hop{{\mathcal H}}
\def\Jop{{\mathcal J}}

\def\Iop{{\mathcal I}}

\def\Iop{{\mathcal I}}

\def\mk{\mathfrak}
\def\txtint{{\textstyle\int}}
\def\wtil{\widetilde}
\def\what{\widehat}
\def\const{\text{const}}

\def\bmatr{\begin{pmatrix}}
\def\ematr{\end{pmatrix}}

\newtheorem{thm}{Theorem}{\bf}{\em}
{\bf}{\em}
\newtheorem{prop}{Proposition}{\bf}{\em}
\newtheorem{lem}{Lemma}{\bf}{\em}
{\bf}{\em}

\def\eqref#1{(\ref{#1})}
\def\Propref#1{Proposition~\ref{#1}}
\def\Thmref#1{Theorem~\ref{#1}}
\def\Thmrefs#1#2{Theorem~\ref{#1} and~\ref{#2}}
\def\Lemref#1{Lemma~\ref{#1}}
\def\Ref#1{Ref.\cite{#1}}
\def\Secref#1{Sec.~\ref{#1}}

\def\ie/{i.e.}
\def\eg/{e.g.}

\def\scrpt#1{$\scriptstyle {#1}$}

\begin{document}

\title{Elastic null curve flows,\\ nonlinear $C$-integrable systems,\\ and geometric realization of Cole-Hopf transformations}

\author{
Z\"uhal K\"u\c{c}\"ukarslan Y\"uzba\c{s}\i${}^1$
\lowercase{and}
Stephen C. Anco$^2$
\\
\\
\lowercase{\scshape{
${}^1$
Department of Mathematics\\
Firat University\\
23119 Elazig, Turkey\\
zuhal\scrpt{2387}@yahoo.com.tr
}}\\
\lowercase{\scshape{
${}^2$
Department of Mathematics and Statistics\\
Brock University\\
St. Catharines, ON L\scrpt2S\scrpt3A\scrpt1, Canada\\
sanco@brocku.ca}}
}

\thanks{sanco@brocku.ca, zuhal2387@yahoo.com.tr.}

\begin{abstract}
Elastic (stretching) flows of null curves are studied in three-dimensional Minkowski space. 
As a main tool, a natural type of moving frame for null curves is introduced,
without use of the pseudo-arclength. 
This new frame is related to a Frenet null frame by a gauge transformation 
that belongs to the little group contained in the Lorentz group $SO(2,1)$
and provides an analog of the Hasimoto transformation 
(relating a parallel frame to a Frenet frame for curves in Euclidean space). 
The Cartan structure equations of the transformed frame are shown to 
encode a hereditary recursion operator giving a two-component generalization of 
the recursion operator of Burgers equation,
as well as a generalization of the Cole-Hopf transformation. 
Three different hierarchies of integrable systems are obtained from the various symmetries of this recursion operator.  
The first hierarchy contains two-component Burgers-type and nonlinear Airy-type systems;
the second hierarchy contains novel quasilinear Schr\"odinger-type (NLS) systems;
and the third hierarchy contains semilinear wave equations (in two-component system form).
Each of these integrable systems is shown to correspond to a geometrical flow of 
a family of elastic null curves in three-dimensional Minkowski space. 
\end{abstract}

\maketitle

\section{Introduction}

The study of geometrical flows of curves has a deep connection with integrable systems. 
One of the simplest examples illustrating this connection 
arises from vortex motion in two and three dimensional inviscid fluid dynamics. 

In two dimensions, 
a patch of constant vorticity in an inviscid fluid can be described by 
the dynamical evolution of its boundary curve. 
This evolution is governed by the vortex boundary-patch equation \cite{GolPet1991,GolPet1992,WexDor}
which is an inelastic (non-stretching) geometrical flow 
$\pos_t = \tfrac{1}{2}\kappa^2\T + \kappa_x\N$ 
for an arclength-parameterized curve $\pos(x,t)$, 
where $\kappa$ is the curvature invariant, $x$ is the arclength, 
and $(\T,\N)$ is the unit tangent vector and unit normal vector along the curve. 
Under this flow, 
the curvature invariant evolves by 
the focusing mKdV equation 
$\kappa_t = \kappa_{xxx} + \tfrac{3}{2}\kappa^2\kappa_x$,
which is an integral system. 

In three dimensions, 
a vortex filament in an inviscid fluid can be described by 
the dynamical evolution of its filament curve,
which is given by the bi-normal equation \cite{Has} 
$\pos_t = \kappa \B$ for an arclength-parameterized curve $\pos(x,t)$. 
A closely related geometrical flow is given by 
$\pos_t = \tfrac{1}{2}\kappa^2\T + \kappa_x\N +\kappa\tau\B$, 
which describes the axial behaviour of a vortex filament \cite{FukMiy}. 
The vortex filament equation and the axial vortex equation are inelastic geometrical flows,
where $(\kappa,\tau)$ are the curvature and torsion invariants, $x$ is the arclength, 
$(\T,\N,\B)$ is the unit tangent vector, unit normal vector, and unit bi-normal vector 
along the curve. 
The evolution of the invariants $(\kappa,\tau)$ of the curve 
under each of the flows yields an integrable system.
These two integrable systems can be transformed \cite{Has,FukMiy,Lam,MarSanWan,AncMyr}
respectively into the focusing NLS equation 
$-iu_t = u_{xx} + \tfrac{1}{2}|u|^2u$ 
and the focusing complex mKdV equation 
$u_t = u_{xxx} + \tfrac{3}{2}|u|^2u_x$
in terms of the variable $u=\kappa \exp(i\int \tau dx)$.
This change of variables, known as the Hasimoto transformation \cite{Has}, 
corresponds geometrically to transforming from the Frenet frame $(\T,\N,\B)$ 
to a parallel frame \cite{Bis} defined by a rotation of the vectors $(\N,\B)$ 
in the normal plane of the curve 
by an angle $\int\tau dx$ at each point on the curve. 
In a parallel frame, $u$ appears as the component in the Cartan matrix 
which describes the infinitesimal transport of the frame along the curve, 
similarly to how $(\kappa,\tau)$ appear as the components in the Cartan matrix 
describing the infinitesimal transport of the Frenet frame $(\T,\N,\B)$. 
Geometrically, the Cartan matrix of a parallel frame represents 
a rotation of the frame vectors around an axis in the normal plane of the curve. 

In both two and three dimensions, 
the integrability structure of these integrable systems 
turns out to be encoded in an explicit way \cite{MarSanWan,AncMyr} 
in the Serret-Frenet equations of the frame formulation of the geometrical flows. 

These well-known results connecting inelastic geometrical curve flows and integrable systems
have been extended to two and three dimensional Minkowski space 
in recent work \cite{AncAlk} on timelike and spacelike curve flows. 

A particularly interesting question is the study flows of null curves in Minkowski space. 
Null curves have qualitatively different features compared to timelike or spacelike curves,
and they arise naturally in many areas of physics, 
such as 
the motion of massless particles and waves (e.g. \cite{PenRin}), 
transport of polarization vectors (e.g. \cite{JosNit}),
the theory of relativistic strings (e.g. \cite{Bun,HugSha}),  
geometrical models of spinning particles (e.g. \cite{NerRam,FerGimLuc2002}). 
They also are of strong interest in mathematics 
and encompass a rich source of geometrical problems
(e.g. \cite{Syn,ONei,FerGimLuc,NurKarTun,DugJin}). 

In contrast to timelike or spacelike curves, 
a main feature of null curves is that their Lorentzian arclength vanishes,
which means that there is no direct analog of inelastic flows. 
One approach which has been used is to consider the pseudo-arclength \cite{Bon} 
defined in terms of the principal normal vector in a Frenet frame
along the null curve. 
Pseudo-inelastic flows in which the pseudo-arclength of the null curve is preserved 
can then be studied \cite{MusNic,AndMar}. 
A completely different approach is to consider elastic flows 
in which the pseudo-arclength of the null curve is allowed to evolve dynamically under the flow. 
Elastic flows have been mainly studied to-date for curves in Euclidean spaces,
such as mean-curvature flow \cite{Gag},
which lead to nonlinear systems that are not integrable. 

A first step in studying elastic flows of null curves was taken in \Ref{AncAlk},
where null curve flows in the two-dimensional Minkowski plane 
were shown to yield Burger's equation 
together with the Cole-Hopf transformation \cite{Col,Hop}
under which Burger's equation is mapped into the heat equation. 
The null curves were formulated by fixing an arbitrary parameterization,
without use of the pseudo-arclength. 
In that setting, 
the dynamical variable in Burger's equation appears as the component of 
the Cartan matrix of a null frame for the parameterized null curves,
and the corresponding geometrical flows consist of an elastic motion 
in which the null tangent vector of the curve 
(and the pseudo-arclength given by the principal normal vector) 
stretches and compresses. 
Moreover, the Serret-Frenet equations of the null frame turn out to encode the Cole-Hopf transformation, 
thereby providing a new geometrical realization for this important transformation
as well as for Burger's equation. 

We will extend this work on elastic null curve flows to three-dimensional Minkowski space
to obtain several main new results in the present paper.
In going from two to three dimensions, 
firstly,
we will introduce a null-vector version of a parallel frame 
that can be applied to arbitrary flows of elastic and inelastic null curves in three-dimensional Minkowski space. 
The gauge transformation relating this new frame to a Frenet null frame 
provides a novel analog of the Hasimoto transformation.
Secondly, 
we will show that the Serret-Frenet equations of the new frame 
encode a two-component hereditary recursion operator 
as well as a two-component Cole-Hopf transformation. 
Thirdly, 
from the various symmetries of this structure, 
we will derive three hierarchies of two-component integrable nonlinear systems:
the first hierarchy contains two-component Burgers-type and nonlinear Airy-type systems;
the second hierarchy contains quasilinear Schr\"odinger-type (NLS) systems;
and the third hierarchy contains semilinear wave equations (in two-component system form). 
The integrability structure of each of the integrable systems consists of 
a hierarchy of higher symmetries and a Cole-Hopf transformation into a linear system. 
In the terminology of Calagero \cite{Cal}, these are $C$-integrable systems. 
We will also show that the Burgers-type system has a gradient flow structure, 
whereas the nonlinear Airy-type system, quasilinear NLS-type system,
and semilinear wave system each have a bi-Hamiltonian structure. 
Finally, 
we will derive the geometrical null curve flows that correspond to these integrable systems. 
These flows describe an elastic motion 
in which the null tangent vector of the curve stretches and compresses, 
while also undergoing a null-rotation,
whereby the pseudo-arclength is time dependent. 

Our results show that the deep connection between nonlinear integrable systems and geometrical inelastic curve flows in Euclidean space
has a natural counterpart for elastic null curve flows in Minkowski space. 

The rest of the paper is organized as follows. 

In \Secref{sec:frames}, 
we first review the construction of a Frenet null frame and its Cartan matrix 
for null curves that have an arbitrary parameterization in $3$-dimensional Minkowski space. 
Next we work out the explicit form of gauge transformations that preserve 
the null tangent vector in this type of Frenet null frame. 
These gauge transformations give a representation of the little group in $SO(2,1)$, 
which we use to generate a general null-tangent frame. 
We then derive the Serret-Frenet equations for this general null-tangent frame
applied to a general elastic null curve flow,
and we show how these equations represent flow equations 
on the components of the Cartan matrix of the general null-tangent frame.

In \Secref{sec:results},
we find a geometrical gauge condition that selects a specific choice of a null-tangent frame
for which the flow equations satisfied by the Cartan matrix components
have a natural formulation involving two potentials, 
leading to a Cole-Hopf transformation that yields 
a two-component linear potential system of decoupled evolution equations. 
This linear system has a natural recursion operator, 
which we use to obtain a corresponding recursion operator for the flow equations
on the Cartan matrix components. 
Next we show how the various symmetries of this recursion operator 
give rise to the three different hierarchies of two-component $C$-integrable nonlinear systems. 
We also discuss the integrability properties of these systems. 
In particular, 
the Burgers-type integrable systems are shown to possess a gradient-flow structure,
while the nonlinear Airy-type integrable systems and the NLS-type integrable systems
are shown to have a bi-Hamiltonian structure. 

In \Secref{sec:gauge}, 
we discuss the properties of the null-tangent frame defined by our geometrical gauge choice.
We work out the explicit form of the little-group gauge transformation 
that relates this frame to a Frenet null frame for null curves 
with an arbitrary parameterization, 
and we show how this describes a novel type of Hasimoto transformation
under which the curvature and torsion variables defined by the Frenet null frame 
are mapped into variables that represent covariants with respect to the little group. 
We also compare this structure to the Hasimoto transformation 
from a Frenet frame to a parallel frame for arclength-parameterized curves 
in Euclidean space. 

In \Secref{sec:flows}, 
we give the derivation of geometrical elastic null curve flows
from the three hierarchies of two-component nonlinear integrable systems. 
We show how these geometrical flows inherit a natural integrability structure,
and we explain more precisely how these flows are elastic in the sense that 
the pseudo-arclength of the null curve dynamically evolves under the flow. 

Finally, we make some concluding remarks in \Secref{sec:conclude}
and discuss some related work on geometrical inelastic (non-stretching) flows of null curves. 
An appendix summarizes the basic structure of Minkowski space that will be used throughout the paper.

\section{Null curve flows in $\Mink{2}$}
\label{sec:frames}

A {\em null curve} in three-dimensional Minkowski space $\Mink{2}$ 
is a curve $\pos(x)$ whose tangent vector $\pos_x$ is null 
\begin{equation}\label{null}
\eta(\pos_{x},\pos_{x})= 0
\end{equation}
at every point $x$ on the curve.
This property does not depend on the choice of parameterization $x$ of the null curve. 

Hereafter we will assume that the span of $\pos_x,\pos_{xx},\pos_{xxx}$ is $3$-dimensional. 
Such null curves will be called {\em non-degenerate}. 
As discussed later, 
this non-degeneracy condition is analogous to the condition in Euclidean space 
that a curve has non-vanishing curvature and torsion.

\subsection{Frenet frames for null curves}
\label{sec:3dimfrenetframe}

For any non-degenerate null curve $\pos(x)$, 
a Frenet-type frame \cite{DugJin} 
can be defined by starting from the null tangent vector 
\begin{equation}\label{T}
\T=\pos_x . 
\end{equation}
Then $\T_x$ defines the principal normal vector 
which satisfies $\eta(\T,\T_x)=0$. 
This relation implies $\T_x$ belongs to the perp space of $\T$, 
and therefore $\T_x$ is a spacelike vector by the null-vector \Lemref{lem2:nullvec}
(see the Appendix)
combined with the non-degeneracy of $\pos(x)$. 
Thus,  
\begin{equation}
\eta(\T_x,\T_x) > 0
\end{equation}
holds at every point $x$ on the curve,
and consequently a unit spacelike normal vector is given by 
\begin{equation}\label{N}
\N=(1/\kappa)\T_x
\end{equation}
with 
\begin{equation}\label{curvinv}
\kappa = \eta(\T_x,\N)= \sqrt{\eta(\T_x,\T_x)} >0 . 
\end{equation}
To complete the frame, 
a vector $\B$ is needed 
that is linearly independent of $\T$ and $\N$ 
and that obeys the normalization $\epsilon(\T,\N,\B)=-1$
where the negative sign is due to the Minkowski signature. 
In Euclidean space, 
this is accomplished by using the cross-product of $\T$ and $\N$,
which yields the bi-normal vector.
However, in Minkowski space, 
$\T\times\N$ is parallel to $\T$,
as shown by the following argument.  
The norm of this cross-product is given by 
$\eta(\T\times\N,\T\times\N) =\eta(\T,\N)^2-\eta(\T,\T)\eta(\N,\N)=0$ 
(see \eqref{crossprod-prop1}), 
which implies $\T\times\N$ is a null vector. 
Since $\T\times\N$ satisfies $\eta(\T\times\N,\N) =0$, 
it is orthogonal to $\N$ which is a spacelike vector. 
Hence we conclude that $\T\times\N$ is a null vector in the Minkowski plane orthogonal to $\N$. 
But since $\T\times\N$ also satisfies $\eta(\T\times\N,\T) =0$, 
it must be proportional to $\T$. 
In particular, by the cross-product property 
$(\T\times\N)\times\N =\eta(\N,\N)\T- \eta(\T,\N)\N$
(see \eqref{crossprod-prop2})
combined with $\eta(\N,\N)=1$ and $\eta(\T,\N)=0$, 
we have $\T\times\N =\T$. 
Consequently, there is no bi-normal vector for null curves in Minkowski space. 

There is, nevertheless, a geometrical way to obtain the needed vector $\B$. 
Since $\T$ is a null vector, 
we can choose $\B$ to be a null vector on the opposite side of the lightcone relative to $\T$ 
in the Minkowski plane orthogonal to $\N$. 
This determines $\B$ up to a multiplicative constant.
The normalization condition 
\begin{equation}\label{orientation}
-1=\epsilon(\T,\N,\B)=\eta(\T,\B)
\end{equation}
then fixes $\B$ uniquely. 

The triple of vectors 
\begin{equation}\label{Frenetframe}
(\T,\N,\B)
\end{equation}
thereby provides a well-defined moving frame along the null curve $\pos(x)$,
with the frame vectors obeying the orthonormality relations 
\begin{gather}
\eta(\T,\T)=\eta(\B,\B)=0,
\quad
\eta(\N,\N)=1,
\label{Frenetframe-rel1}
\\
\eta(\T,\N)=\eta(\N,\B)=0,
\quad
\eta(\T,\B)=-1 . 
\label{Frenetframe-rel2}
\end{gather}
This is called a {\em Frenet null frame} \cite{DugJin}
with respect to an arbitrary smooth parameterization of the null curve $\pos(x)$. 
Null frames have been important in previous work on null curve flows 
in the Minkowski plane \cite{AncAlk}, 
especially for deriving integrable systems from the frame structure equations. 

The Serret-Frenet equations of this frame \eqref{Frenetframe}
are straightforward to derive. 
First, from equation \eqref{N}, we have
\begin{equation}\label{Tx}
\T_x=\kappa \N . 
\end{equation}
Next, we can write $\N_x=a\T+b\N+c\B$
for some functions $a(x),b(x),c(x)$ which are determined 
using relations \eqref{Frenetframe-rel1}--\eqref{Frenetframe-rel2}
as follows
From $0=\eta(\N,\N_x)= b$, we find $b=0$,
while from $\eta(\T,\N_x)= -c$ and $\eta(\T_x,\N)=\kappa$, 
we get $c=\kappa$. 
Note we have $\eta(\B,\N_x) =-a$. 
In analogy with a Euclidean Frenet frame, 
we write 
\begin{equation}\label{torsinv}
\tau = \eta(\N_x,\B) ,
\end{equation}
and so we have $a=-\tau$,
which gives 
\begin{equation}\label{Nx}
\N_x= -\tau\T +\kappa\B . 
\end{equation}
Finally, we can write $\B_x=\tilde a\T+\tilde b\N+\tilde c\B$
and proceed in a similar way to determine the functions $\tilde a(x),\tilde b(x),\tilde c(x)$. 
This yields
\begin{equation}\label{Bx}
\B_x= -\tau\N . 
\end{equation}

We note that the the non-degeneracy condition on the curve
is equivalent to $\kappa$ and $\tau$ both being non-zero. 
We also note that equation \eqref{Nx} for the normal vector $\N$
provides an alternative geometrical way to define the null vector $\B$ in the Frenet null frame. 

Thus, the Serret-Frenet equations \eqref{Tx}, \eqref{Nx}, \eqref{Bx}
are given by
\begin{equation}\label{Freneteqns}
\bmatr 
\T_x\\  \N_x \\ \B_x
\ematr
=\bmatr 
0 & \kappa & 0\\ 
-\tau & 0 & \kappa \\ 
0  & -\tau & 0 
\ematr
\bmatr 
\T\\  \N \\ \B
\ematr
\end{equation}
where the $3\times3$ matrix belongs to a representation of the Lie algebra 
$\mk{so}(2,1)$ of the $SO(2,1)$ Lorentz group of isometries in $\Mink{2}$.
This frame is the natural Lorentzian counterpart of a Frenet frame,
adapted to null curves with an arbitrary smooth parameterization. 

There are several key differences between this Frenet null frame for null curves in $\Mink{2}$ 
and a Frenet frame for curves in Euclidean space $\Rnum^3$. 

First, Euclidean curves $\pos(x)$ have three invariants, 
consisting of the arclength, curvature, and torsion,
with arclength being defined by $ds = \sqrt{\pos_x\cdot\pos_x}dx$. 
For null curves, the analogous arclength is trivial since $\pos_x$ is null. 

Second, the Euclidean curvature and torsion are defined in terms of a Frenet frame 
using an arclength parameterization:
$\kappa = \eta(\T_s,\N)$ and $\tau = \eta(\N_s,\B)$,
with $(\T,\N,\B)$ being the unit tangent vector, unit normal vector, and unit bi-normal vector, respectively. 
Since the arclength is an invariant, 
so are the curvature $\kappa(s)$ and the torsion $\tau(s)$. 
In contrast, 
the corresponding quantities \eqref{curvinv} and \eqref{torsinv} for null curves 
are defined relative to an arbitrary parameterization $x$,
and so $\kappa(x)$ and $\tau(x)$ are not invariants.
Specifically, under a smooth reparameterization $x\to \tilde x=f(x)$ with $f'(x)>0$, 
we see that 
\begin{equation}\label{reparam}
\kappa(x)\to \tilde\kappa(\tilde x)= \kappa(x)/f'(x)^2,
\quad
\tau(x)\to \tilde\tau(\tilde x) = (\tau(x)+2\sqrt{1/f'(x)}((\sqrt{f'(x)})_x/\kappa(x))_x)/f'(x) .
\end{equation}
We will call $\kappa(x)$ and $\tau(x)$ 
the \emph{Frenet curvature} and \emph{Frenet torsion}, respectively. 
The natural invariants for null curves are instead given by \cite{Bon} 
the pseudo-arclength $ds = \eta(\T_x,\T_x)^{1/4}dx$ 
and the null curvature $k = \eta(\N_s,\B)$, 
with $\T=\pos_s$ being the null tangent vector, 
$\N=\T_s$ being the principal normal vector, and $\B$ being the null opposite vector relative to $\T$ and $\N$. 
We will not use the pseudo-arclength parameterization of null curves here
because we will be interested in studying elastic flows that do not preserve this arclength.

Third, for null curves, 
neither of the null vectors $\T$ and $\B$ can be obtained 
from cross-products of the other two frame vectors. 
In particular, the cross-product yields 
$\T\times\N =\T$, $\B\times\N =-\B$, 
while in contrast, $\T\times\B=\N$. 
The geometrical meaning of the pair $\B$ and $\T$ is they 
can be viewed as ingoing and outgoing null vectors 
in the Minkowski plane orthogonal to $\N$. 

For notational clarity, 
hereafter we will write 
\begin{equation}\label{Frenetnullframe}
\e_{+}=\T,
\quad
\e_{-}=\B,
\quad
\e_{\perp} = \N
\end{equation}
whereby the geometrical relations \eqref{Frenetframe-rel1}--\eqref{Frenetframe-rel2} and \eqref{orientation}
on $\T,\N,\B$ become 
\begin{gather}
\eta(\e_{+},\e_{+})=\eta(\e_{-},\e_{-})=0,
\quad
\eta(\e_{+},\e_{-})=-1,
\label{Frenetnullframe-rel1}
\\
\eta(\e_{\pm},\e_{\perp})=0,
\quad
\eta(\e_{\perp},\e_{\perp})=1,
\label{Frenetnullframe-rel2}
\\
\epsilon(\e_{+},\e_{-},\e_{\perp})=1 . 
\end{gather}
We note that, geometrically, 
$\tfrac{1}{\sqrt{2}}(\e_{+}+\e_{-})$ is a timelike unit vector
which is invariant under spatial reflections that interchange the two null vectors 
$\e_{+}$ and $\e_{-}$
in the Minkowski plane orthogonal to $\e_{\perp}$. 

We will now summarize this construction. 

\begin{prop}
Any non-degenerate null curve in $\Mink{2}$ 
possesses a well-defined Frenet null frame
\eqref{Frenetnullframe}--\eqref{Frenetnullframe-rel2}
in terms of the null tangent vector \eqref{T},
using an arbitrary smooth parameterization. 
The Serret-Frenet equations of this frame 
\begin{equation}\label{nullFrenetframe}
\E = \bmatr 
\e_{+} \\ \e_{-} \\ \e_{\perp} 
\ematr
\end{equation}
consist of 
\begin{equation}\label{nullFreneteqns}
\E_x = \U \E
\end{equation}
given by the Cartan matrix 
\begin{equation}\label{nullFrenetU}
\U=
\bmatr 
0 & 0 & \kappa \\ 
0 & 0 & -\tau \\ 
-\tau & \kappa & 0 
\ematr 
\end{equation}
which belongs to the Lie algebra $\mk{so}(2,1)$ of the $SO(2,1)$ group of 
rotation and boost isometries in $\Mink{2}$.
The components $(\kappa,\tau)$ of this matrix define the Frenet curvature and torsion 
(cf \eqref{curvinv}, \eqref{torsinv}) of the curve. 
\end{prop}

\subsection{Gauge transformations}

A general moving frame for a null curve $\pos(x)$ in $\Mink{2}$ is related to 
the Frenet null frame \eqref{nullFrenetframe}
by the action of an $x$-dependent $SO(2,1)$ transformation group 
on the frame vectors $(\e_{+},\e_{-},\e_{\perp})$. 
If the tangent vector $\T=\e_{+}$ is preserved as one of the frame vectors, 
then the appropriate transformation group will be a subgroup representation of 
the little group in $SO(2,1)$. 
This subgroup is generated by an infinitesimal transformation
\begin{equation}
\E\to \wtil\E = \E + \epsilon S\E + O(\epsilon^2)
\end{equation}
with parameter $\epsilon(x)$, 
where $S$ is a $3\times3$ constant matrix 
(in a null-vector representation of $\mk{so}(2,1)$)
determined by the condition that the orthonormality relations \eqref{Frenetnullframe-rel1}--\eqref{Frenetnullframe-rel2} 
are preserved to $O(\epsilon^2)$. 
A straightforward computation, which will be omitted, yields
\begin{equation}
S = \bmatr 
0 & 0 & 0 \\ 
0 & 0 & 1 \\
1 & 0 & 0
\ematr . 
\end{equation}
This matrix generates the transformation group
\begin{equation}\label{gaugegroup}
\S(\epsilon) = \exp(\epsilon S)
= \bmatr 
1 & 0 & 0 \\ 
\tfrac{1}{2}\epsilon^2 & 1 & \epsilon \\
\epsilon & 0 & 1
\ematr
\end{equation}
comprising a representation of the little group in $SO(2,1)$. 
Note that the group parameter $\epsilon(x)$ can be an arbitrary function of $x$ along the curve. 
Then the action of $\S(\epsilon)$ on the Frenet null frame is given by 
\begin{equation}\label{SonE}
\E\to \wtil\E = \S(\epsilon)\E
\end{equation}
which yields the transformed frame vectors 
\begin{equation}\label{nulltangentframe}
\wtil\E = \bmatr 
\tilde\e_{+} \\ \tilde\e_{-} \\ \tilde\e_{\perp} 
\ematr,
\quad
\tilde\e_{+} = \e_{+}, 
\quad
\tilde\e_{-} = \e_{-} +\epsilon \e_{\perp} +\tfrac{1}{2} \epsilon^2 \e_{+},
\quad
\tilde\e_{\perp} = \e_{\perp} + \epsilon \e_{+}
\end{equation}
satisfying
\begin{equation}\label{nulltangentframe-rels}
\eta(\tilde\e_{+},\tilde\e_{+})=\eta(\tilde\e_{-},\tilde\e_{-})=0,
\quad
\eta(\tilde\e_{\pm},\tilde\e_{\perp})=0,
\quad
\eta(\tilde\e_{+},\tilde\e_{-})=-1,
\quad
\eta(\tilde\e_{\perp},\tilde\e_{\perp})= 1.
\end{equation}
The Serret-Frenet matrix equation \eqref{nullFreneteqns} becomes 
\begin{equation}\label{nulltangentFreneteqns}
\wtil\E_x = \wtil\U\wtil\E
\end{equation}
where the transformed Cartan matrix 
\begin{equation}
\U\to \wtil\U = (\S(\epsilon)_x+\S(\epsilon)\U)\S(\epsilon)^{-1}
\end{equation}
is given by 
\begin{equation}\label{nulltangentU}
\wtil\U=
\bmatr 
u_0 & 0 & u_{+} \\ 
0 & -u_0 & u_{-} \\ 
u_{-} & u_{+} & 0 
\ematr
\end{equation}
with 
\begin{equation}\label{Ucomps}
u_0 = -\epsilon\kappa,
\quad
u_{+} = \kappa,
\quad
u_{-} = -\tau -\tfrac{1}{2}\epsilon^2\kappa +\epsilon_x .
\end{equation}
Note, when $\epsilon\neq 0$, 
the form of $\wtil\U$ algebraically differs from the form of $\U$ 
by having a non-vanishing component $u_0$. 
This shows that the transformed frame is no longer a Frenet null frame.

\subsection{Elastic flow equations}

We now consider null curve flows $\pos(x,t)$ that locally preserve 
both the null signature \eqref{null} and the fixed parameterization $x$ of the curve.
Such flows are specified by the vector 
\begin{equation}\label{flow}
\pos_t=h_{+}\tilde\e_{+}+h_{-}\tilde\e_{-} +h_{\perp}\tilde \e_{\perp}
\end{equation}
expressed in terms of the frame vectors in a general null-tangent frame \eqref{nulltangentframe} 
at each point $x$ along the curve. 

Since the tangent vector $\pos_x=\tilde\e_{+}$ remains null under these flows, 
there will be an induced flow on the null-tangent frame, 
such that the normalization relations \eqref{nulltangentframe-rels} 
are preserved. 
This implies that the $t$-derivative of the frame vectors \eqref{nulltangentframe}
is given by 
\begin{equation}\label{nulltangentfloweqns}
\wtil\E_t=\wtil\W\wtil \E
\end{equation}
where 
\begin{equation}\label{nulltangentW}
\wtil\W=
\bmatr 
\omega_0 & 0 & \omega_{+} \\ 
0 & -\omega_0 & \omega_{-} \\ 
\omega_{-} & \omega_{+} & 0 
\ematr
\end{equation}
is the Cartan flow matrix, 
which belongs to the Lie algebra $\mk{so}(2,1)$. 
From the form of this matrix \eqref{nulltangentW}, 
note $\eta(\pos_x,\pos_x)_t = 2 \eta(\tilde\e_{+},\tilde\e_{+t}) = 2 \omega_0\eta(\tilde\e_{+},\tilde\e_{+}) +2\omega_{+}\eta(\tilde\e_{+},\tilde\e_{\perp}) =0$ holds,
which is necessary and sufficient for the flow \eqref{flow} 
to preserve the null signature \eqref{null} of the curve. 

This flow equation \eqref{nulltangentfloweqns} on the frame
and the Serret-Frenet equation \eqref{nullFreneteqns} of the frame 
are related by the compatibility condition 
$\partial_t(\wtil\E_x)=\partial_x(\wtil\E_t)$. 
This condition is equivalent to a zero curvature equation
\begin{equation}
\wtil\U_t-\wtil\W_x+[\wtil\U,\wtil\W]=0
\label{zerocurv}
\end{equation}
relating the Cartan matrices $\wtil\W$ and $\wtil\U$. 
After substituting these matrices \eqref{nulltangentU} and \eqref{nulltangentW}
into equation \eqref{zerocurv}, 
we obtain
\begin{align}
& u_{+t}-\omega_{+x} -u_{+}\omega_0 +u_0\omega_{+} =0, 
\label{u+t}\\
& u_{-t} -\omega_{-x} +u_{-}\omega_0 -u_0\omega_{-} =0, 
\label{u-t}\\
& u_{0t} -\omega_{0x} +u_{+}\omega_{-} -u_{-}\omega_{+} =0.
\label{u0t}
\end{align}

Likewise, 
the flow vector \eqref{flow}
and the null tangent vector \eqref{T}
of the curve are related by the compatibility condition 
$\partial_x(\pos_t)=\partial_t(\pos_x)$. 
We write 
\begin{equation}
\pos_x=\A^\t\wtil\E,
\quad
\pos_t=\wtil\H^\t\wtil\E
\end{equation}
where 
\begin{equation}\label{HA}
\wtil\H=\bmatr h_{+} \\ h_{-} \\ h_{\perp} \ematr, 
\quad
\A= \bmatr 1 \\ 0 \\ 0 \ematr
\end{equation}
are column vectors giving the components of the null tangent vector and the flow vector
in terms of the null-tangent frame,
with $\A$ being invariant under the transformations \eqref{gaugegroup} 
representing the little group. 
Then the compatibility condition becomes 
\begin{equation}
\wtil\H_x+\wtil\U^\t\wtil\H = \wtil\W^\t\A
\label{zerotors}
\end{equation}
relating $\wtil\W$ to $\wtil\U$ and $\wtil\H$. 
After we substitute the Cartan matrices \eqref{nulltangentU} and \eqref{nulltangentW}
along with the vectors \eqref{HA} 
into equation \eqref{zerotors}, 
we find 
\begin{align}
& h_{+x} -\omega_0 + u_{-} h_{\perp} +u_0 h_{+} =0, 
\label{h+x}\\
& h_{-x} + u_{+} h_{\perp} - u_0 h_{-} =0, 
\label{h-x}\\
& h_{\perp x} -\omega_{+} + u_{+} h_{+} + u_{-} h_{-} =0.
\label{hperpx}
\end{align}

Taken together, 
the compatibility equations \eqref{h+x}--\eqref{hperpx} and \eqref{u+t}--\eqref{u0t} 
give a way to formulate all null curve flows $\pos(x,t)$ in $\Mink{2}$.
In particular, 
if the flow components $(h_{+},h_{\perp})$ and $\omega_{-}$ are specified 
as arbitrary functions of the parameter $x$ along the curve for all $t\geq 0$, 
then $\omega_0,h_{-},\omega_{+}$ are respectively determined by equations \eqref{h+x}--\eqref{hperpx},
while $(u_{+},u_{-},u_0)$ evolve by equations \eqref{u+t}--\eqref{u0t}. 

This establishes the following general result. 

\begin{prop}\label{nullcurveflows}
All flows of non-degenerate null curves $\pos(x,t)$ in $\Mink{2}$
are described by the system of equations \eqref{u+t}--\eqref{u0t} and \eqref{h+x}--\eqref{hperpx}
formulated using a general null-tangent frame \eqref{nulltangentframe}. 
In this system, 
the components $(u_{+},u_{-},u_0)$ of the Cartan matrix 
are related to the Frenet curvature and torsion $(\kappa,\tau)$ of the null curve 
by expression \eqref{Ucomps}
involving a freely specifiable gauge parameter $\epsilon(x)$. 
\end{prop}

\section{Integrable systems in $\Mink{2}$}
\label{sec:results}

The flow equations derived in \Propref{nullcurveflows} 
for parameterized null curves $\pos(x,t)$ in $\Mink{2}$
have a gauge freedom coming from the use of a general null-tangent frame \eqref{nulltangentframe}
which is related to a Frenet null frame \eqref{nullFrenetframe}
by the little group of gauge transformations \eqref{gaugegroup}. 
Our aim is to seek a gauge choice under which these flow equations
reduce to a two-component evolution system that encodes a hereditary recursion operator. 

We start by considering the special case when a null curve flow $\pos(x,t)$ 
is constrained to a Minkowski plane $\Mink{1} \subset \Mink{2}$,
since the flow equations are then known to encode 
the hereditary recursion operator of Burgers equation,
as shown in \Ref{AncAlk}. 
This constraint is given by $h_0=0$ 
in terms of the components $(h_{+},h_{-},h_0)$ 
in the flow vector \eqref{flow}, 
and $u_{\pm}=\omega_{\pm}=0$ 
in terms of the components $(u_{+},u_{-},u_0)$, $(\omega_{+},\omega_{-},\omega_0)$
in the Cartan matrices \eqref{nulltangentU} and \eqref{nulltangentW}. 
The flow equations \eqref{u+t}--\eqref{u0t} and \eqref{h+x}--\eqref{hperpx} 
given by \Propref{nullcurveflows} 
thereby become 
\begin{equation}\label{2dimu0tw0t}
u_{0t} = \omega_{0x},
\quad
\omega_0 = h_{+x} + u_0 h_{+}
\end{equation}
and
\begin{equation}\label{2dimh-x}
h_{-x} - u_0 h_{-} =0 . 
\end{equation}
The first two equations can be expressed in a potential form 
$u_0 = v_{0x}$ and $\omega_0 = v_{0t} = h_{+x} + v_{0x} h_{+}$,
yielding the one-component system 
\begin{equation}\label{2dimpotential}
v_{0t} = (D_x + v_{0x}) h_{+} .
\end{equation}
The third equation gives $u_0 = h_{-x}/h_{-}$, 
which yields $v_0=\ln(h_{-})$. 
We then recognize $D_x +v_{0x}$ is the hereditary recursion operator 
for Burgers equation $u_{0t} = u_{0xx} + 2u_0u_{0x}$ in potential form $v_{0t}= v_{0xx} + v_{0x}^2$,
which arises from taking $h_{+}=v_{0x}$. 
Also, we recognize $v_0=\ln(h_{-})$ is 
the Cole-Hopf transformation that maps this form of Burgers equation 
into the linear heat equation $h_{-t}= h_{-xx}$. 

The closest extension of this structure appearing in the full system of flow equations \eqref{u+t}--\eqref{u0t} and \eqref{h+x}--\eqref{hperpx} 
in Minkowski space $\Mink{2}$
arises when $u_{-}=\omega_{-}=0$. 
Specifically, 
we see equations \eqref{u0t} and \eqref{h+x} 
reduce exactly to the pair of equations \eqref{2dimu0tw0t}; 
equations \eqref{u+t} and \eqref{hperpx} 
reduce to a similar pair of equations; 
equation \eqref{h-x} becomes similar to equation \eqref{2dimh-x};
and finally equation \eqref{u-t} becomes trivial. 

Hence, we will hereafter impose the gauge choice 
\begin{equation}\label{gaugecond}
u_{-}=0
\end{equation}
on the Cartan matrix \eqref{nulltangentU}. 
Using expression \eqref{Ucomps} for $(u_{+},u_{-},u_0)$ 
in terms of the gauge parameter $\epsilon(x)$, 
we see that this gauge choice can be achieved by taking 
\begin{equation}\label{gaugerel1}
\epsilon_x = \tau +\tfrac{1}{2}\epsilon^2\kappa
\end{equation}
which is a Riccati equation for $\epsilon(x)$. 
A standard transformation 
\begin{equation}
\epsilon = (-1/\kappa)\big( \ln(\kappa)_x +2\ln(\mu)_x \big)
\end{equation}
can be used to convert the Riccati equation into a linear second-order differential equation
\begin{equation}\label{gaugerel2}
\mu_{xx} + \tfrac{1}{2}\big( \tau\kappa +\kappa_{xx}/\kappa -\tfrac{3}{2}(\kappa_x/\kappa)^2 \big)\mu =0
\end{equation}
for $\mu(x)$. 
Note the solution for $\epsilon$ will be a nonlocal expression 
in terms of the Frenet curvature $\kappa$ and the Frenet torsion $\tau$ of the null curve,
involving an arbitrary constant. 
The presence of this constant represents a residual gauge freedom 
which we will discuss later (see \Secref{sec:gauge}). 
The other two components in the Cartan matrix \eqref{nulltangentU}
are then given by 
\begin{equation}\label{gaugeUcomps}
u_0 = -\epsilon\kappa = \ln(\mu^2\kappa)_x , 
\quad
u_{+} = \kappa .
\end{equation}

In this gauge, 
the flow equations \eqref{u+t}--\eqref{u0t} and \eqref{h+x}--\eqref{hperpx} 
reduce to the system 
\begin{align}
& u_{0t} = \omega_{0x} -u_{+}\omega_{-} , \\
& u_{+t} = \omega_{+x} +u_{+}\omega_0 -u_0\omega_{+} , \\
& \omega_{-x} + u_0\omega_{-} =0 ,
\label{w-}
\end{align}
and
\begin{align}
& \omega_0 = h_{+x} + u_0 h_{+} , \\
& \omega_{+} = h_{\perp x} + u_{+} h_{+} ,\\
& h_{-x} + u_{+} h_{\perp} -u_0 h_{-} =0 .
\end{align}
Observe the component $\omega_{-}$ satisfies 
a linear homogeneous differential equation \eqref{w-} 
which does not involve any components $(h_{+},h_{-},h_0)$ of the flow vector. 
Consequently, without constraining the flow, 
we are free to impose the condition 
\begin{equation}\label{flowcond}
\omega_{-}=0 .
\end{equation}

The resulting system for $(u_{+},u_0)$, $(\omega_{+},\omega_0)$, and $(h_{+},h_{-},h_0)$
is given by 
\begin{align}
& 
u_{0t} = \omega_{0x}, 
\quad
\omega_0 =h_{+x} + u_0 h_{+} ,
\label{gaugeu0eqn}\\
& 
u_{+t} = \omega_{+x} + u_{+}\omega_0 -u_0\omega_{+},
\quad
\omega_{+}  = h_{\perp x}  + u_{+} h_{+} ,
\label{gaugeu+eqn}\\
& 
h_{-x} -u_0 h_{-} +u_{+} h_{\perp} =0 .
\label{gaugeh-eqn}
\end{align}
Note equations \eqref{gaugeu0eqn} and \eqref{gaugeu+eqn}
have the two-component form 
\begin{equation}
u_{0t} = h_{+xx} + (u_0 h_{+})_x , 
\quad
u_{+t} = h_{\perp xx}  + (u_{+} h_{+})_x + u_{+}h_{+x} -u_0h_{\perp x} .
\label{gaugeu0u+sys}
\end{equation}

We expect that this system \eqref{gaugeu0u+sys} and \eqref{gaugeh-eqn}
should encode a hereditary recursion operator 
after a suitable Cole-Hopf transformation. 
Recall, a linear pseudo-differential operator $\Dop$ is hereditary 
iff it satisfies \cite{Olv-book}
\begin{equation}\label{hereditaryprop}
\lieder{\Dop\X}\Dop = \Dop(\lieder{\X}\Dop)
\end{equation}
for all vector fields $\X$, 
where $\lieder{}$ denotes the Lie derivative \cite{Olv-book}. 

To proceed, 
first, we observe that equation \eqref{gaugeu0eqn}
can be expressed in a natural potential form given by 
\begin{equation}\label{u0potential}
u_0=v_{0x},
\quad
\omega_0 = v_{0t} = h_{+x} + v_{0x} h_{+} .
\end{equation}
Next, by use of the integrating factor $e^{-v_0}$,
we see that equation \eqref{gaugeu+eqn} can be combined with equation \eqref{u0potential}
into a similar potential form 
\begin{equation}\label{u+potential}
e^{-v_0}u_{+}=v_{+x},
\quad
e^{-v_0}\omega_{+}= v_{+t} = e^{-v_0} h_{\perp x} + v_{+x} h_{+} .
\end{equation}
In the same way, 
these two potential equations \eqref{u0potential} and \eqref{u+potential} 
can be expressed as 
\begin{equation}\label{V0V+eqn}
V_{0t}= H_{1x},
\quad
V_{+t}= H_{2x} ,
\end{equation}
with 
\begin{align}
& 
V_0=e^{v_0}, 
\quad
V_{+}=e^{v_0}v_{+} ,
\label{v0v+potential}\\
& 
H_1 = e^{v_0}h_{+},
\quad
H_2 = h_{\perp} +e^{v_0}v_{+}h_{+} .
\label{H+Hperp}
\end{align}

Hence, 
we have arrived at a two-component linear evolution system 
in terms of the potentials $(V_0,V_{+})$
which are related to $(u_0,u_{+})$ by the transformations
\begin{equation}\label{colehopf}
\begin{aligned}
& u_0=v_{0x} = V_{0x}/V_0 = \ln(V_0)_x, \\
& u_{+}= e^{v_0} v_{+x} = V_{+x} - V_{+}V_{0x}/V_0 = V_0 (V_{+}/V_0)_x .
\end{aligned}
\end{equation}

The two-component linear evolution system \eqref{V0V+eqn}
has the operator formulation 
\begin{equation}\label{linearsys}
\bmatr
V_0 \\ V_{+}
\ematr_t
= 
D_x \bmatr
H_1 \\ H_2
\ematr
\end{equation}
where $D_x$ is well-known to be both a Hamiltonian operator and a hereditary recursion operator \cite{Olv-book}. 

When $D_x$ is viewed as a hereditary recursion operator, 
then $(H_1,H_2)$ represents the components of a vector field 
\begin{equation}\label{linearsysX}
\X = H_1\partial_{V_0} + H_2\partial_{V_{+}}
\end{equation}
in the jet space $(t,x,V_0,V_{+},V_{0t},V_{+t},V_{0x},V_{+x},\ldots)$
associated to the potentials $(V_0,V_{+})$. 
This structure will be preserved under the inverse of the transformations \eqref{colehopf}. 

Firstly, we go back to the potentials \eqref{v0v+potential}. 
Transforming the vector field \eqref{linearsysX}
by the standard vector transformation rule, 
we get 
\begin{equation}\label{potentialsysX}
\X = h_1\partial_{v_0} + h_2\partial_{v_{+}}
\end{equation}
with 
\begin{equation}\label{potentialsysXcomponents}
h_1= e^{-v_0} H_1, 
\quad
h_2 = e^{-v_0}(H_2 - v_{+} H_1) .
\end{equation}
We next use these components of $\X$ to express 
the two-component potential system \eqref{u0potential}--\eqref{u+potential}
in an operator form
\begin{equation}\label{potentialsys}
\bmatr
v_0 \\ v_{+}
\ematr_t
= 
\bmatr 
D_x +v_{0x} & 0 \\ 
v_{+x} & D_x +v_{0x} 
\ematr
\bmatr
h_1 \\ h_2
\ematr .
\end{equation}
This operator
\begin{equation}\label{potentialRop}
\wtil\Rop = 
\bmatr 
D_x +v_{0x} & 0 \\ 
v_{+x} & D_x +v_{0x} 
\ematr
\end{equation}
can be invertibly mapped into the constant-coefficient diagonal operator
\begin{equation}\label{Dxop}
\what\Rop = \bmatr 1 & 0 \\ 0 & 1 \ematr D_x
\end{equation}
under the transformation \eqref{v0v+potential},
where all of these operators transform as $(1,1)$ tensors \cite{Olv-book}. 
Because the diagonal operator \eqref{Dxop} has constant coefficients, 
it is a hereditary recursion operator \cite{Olv-book}. 
As a consequence, 
$\wtil\Rop$ is also a hereditary recursion operator, 
since the hereditary property \eqref{hereditaryprop} is preserved under invertible transformations. 

Secondly, we go back to the components $(u_0,u_{+})$, 
using the transformation \eqref{colehopf}. 
We find that the vector field \eqref{potentialsysX} transforms into 
\begin{equation}\label{X}
\X = f_1\partial_{u_0} + f_2\partial_{u_{+}}
\end{equation}
with 
\begin{equation}\label{Xcomponents}
f_1= h_{1x}, 
\quad
f_2 = e^{v_0} h_{2x} + u_{+} h_1
\end{equation}
where $v_0=\int u_0\,dx$. 
Using these components $(f_1,f_2)$ of $\X$, 
we express the two-component system \eqref{gaugeu0u+sys} for $(u_0,u_{+})$
in an operator form
\begin{equation}\label{sys}
\bmatr
u_0 \\ u_{+}
\ematr_t
= 
\bmatr 
D_x +D_x u_0D_x^{-1} & 0 \\ 
u_{+}-u_{0x}(D_x -u_0)^{-1}u_{+}D_x^{-1} & (D_x -u_0)D_x(D_x -u_0)^{-1}
\ematr
\bmatr
f_1 \\ f_2
\ematr .
\end{equation}
This nonlocal operator
\begin{equation}\label{Rop}
\Rop = 
\bmatr 
D_x +D_x u_0D_x^{-1} & 0 \\ 
u_{+}-u_{0x}(D_x -u_0)^{-1}u_{+}D_x^{-1} & (D_x -u_0)D_x(D_x -u_0)^{-1}
\ematr
\end{equation}
can be invertibly mapped into the local operator $\wtil\Rop$ 
under the transformation \eqref{colehopf},
and consequently $\Rop$ is a hereditary recursion operator. 
The corresponding transformation of the vector field components is given by 
\begin{equation}\label{colehopfX}
f_1= D_x h_1 = D_x((1/V_0)H_1), 
\
f_2 = e^{v_0}(D_x h_2 + v_{+x} h_1) = V_0D_x((1/V_0)H_2) -V_{+}D_x((1/V_0)H_1) .
\end{equation}
By equating the system \eqref{sys} to the original system \eqref{gaugeu0u+sys} for $(u_0,u_{+})$, 
we obtain a relation directly giving the vector field components $(f_1,f_2)$ 
in terms of the flow vector components $(h_{+},h_{\perp})$:
\begin{equation}
f_1 = D_x h_{+}, 
\quad
f_2 = D_x h_{\perp} -u_0 h_{\perp} + u_{+}h_{+} .
\end{equation}
This relation can be inverted to get
\begin{equation}\label{h+hperp}
h_{+} = D_x^{-1} f_1, 
\quad
h_{\perp} = (D_x -u_0)^{-1}(f_2 -u_{+}D_x^{-1}f_1) .
\end{equation}

Finally,
we note $h_{-}$ can be found in terms of $h_{+}$, $u_0$ and $u_{+}$ 
from equation \eqref{gaugeh-eqn}. 
By applying the integrating factor $e^{-v_0}$ to this equation, 
we get
$(e^{-v_0}h_{-})_x = - e^{-v_0}u_{+} h_{\perp}$,
yielding
\begin{equation}\label{h-}
h_{-} = -(D_x-u_0)^{-1}(u_{+}h_{\perp}) .
\end{equation}

The preceding results can be summarized as follows. 

\begin{thm}\label{structurethm}
All flows of non-degenerate null curves 
(with an arbitrary smooth parameterization $x$) in $\Mink{2}$ 
can be formulated in a null-tangent frame \eqref{nulltangentframe}
satisfying the gauge conditions \eqref{gaugecond} and \eqref{flowcond}. 
The components of the Cartan matrix in this gauge 
are given in terms of the Frenet curvature $\kappa$ and torsion $\tau$ of the null curve 
by 
\begin{equation}\label{u0u+}
u_{+} = \kappa,
\quad
u_0 = -\epsilon(\kappa,\tau)\kappa 
= \ln(\mu(\kappa,\tau)^2\kappa)_x
\end{equation}
where $\epsilon(\kappa,\tau)$ is determined by a Riccati equation \eqref{gaugerel1}
and $\mu(\kappa,\tau)$ is determined by an equivalent linear second-order differential equation \eqref{gaugerel2}. 
These components $(u_0,u_{+})$ satisfy a nonlinear evolution system \eqref{sys} 
involving a hereditary recursion operator \eqref{Rop}
and two freely specifiable functions $(f_1,f_2)$ of $x,t$. 
The flow vector in this formulation is expressed as 
\begin{equation}\label{flowvec}
\pos_t= h_{+}\tilde\e_{+}+h_{-}\tilde\e_{-} +h_{\perp}\tilde\e_{\perp}
\end{equation}
where its components $(h_{+},h_{\perp})$ are given in terms of $(f_1,f_2)$ 
by expression \eqref{h+hperp}, 
while its remaining component is given by expression \eqref{h-}
involving $h_{\perp}$, $u_0$, $u_{+}$. 
Moreover, 
the nonlinear evolution system for  $(u_0,u_{+})$ 
can be mapped into a linear evolution system \eqref{linearsys}
by a Cole-Hopf transformation \eqref{colehopf} and \eqref{colehopfX}, 
under which the hereditary recursion operator \eqref{Rop} is mapped into 
the constant-coefficient diagonal operator \eqref{Dxop}.
\end{thm}

\subsection{Two-component integrable Burgers system}

By a standard result in the theory of recursion operators \cite{Olv-book}, 
any hereditary operator can be used to derive an integrable system 
starting from a symmetry of the operator. 

The hereditary operator \eqref{Rop} in \Thmref{structurethm}
is manifestly invariant under $x$-translations 
which are generated by the symmetry vector field 
$\X=u_{0x}\partial_{u_0} +u_{+x}\partial_{u_{+}}$. 
An integrable system is obtained from this symmetry 
by substitution of its vector field components 
\begin{equation}
f_1=u_{0x},
\quad
f_2=u_{+x}
\end{equation}
into the nonlinear evolution system \eqref{sys}, 
yielding
\begin{equation}\label{Burgerssys}
u_{0t} = u_{0xx} + 2u_0u_{0x},
\quad
u_{+t} = u_{+xx} + 2u_{+}u_{0x} .
\end{equation}
This is a two-component Burgers system. 
It is integrable in the sense that the hereditary recursion operator \eqref{Rop}
generates a hierarchy of higher-order local symmetries starting from the $x$-translation symmetry. 
These symmetries are given by the vector fields
\begin{equation}\label{Burgerssymms}
\X_{n} = f_1^{(n)}\partial_{u_0} +f_2^{(n)}\partial_{u_{+}}
\quad
n=0,1,2,\ldots
\end{equation}
with 
\begin{equation}\label{Burgershierarchy}
\bmatr
f_1^{(n)} \\
f_2^{(n)} 
\ematr
= \Rop^n
\bmatr
u_{0x} \\
u_{+x}
\ematr,
\quad
n=0,1,2,\ldots
\end{equation}
where the first symmetry $\X_0$ in this hierarchy is the $x$-translation symmetry, 
and where the second symmetry $\X_1$ 
corresponds to the Burgers system \eqref{Burgerssys} itself.
The Burgers system also possesses a family of scaling symmetries
\begin{equation}\label{Burgersscalsymm}
t\to \lambda^2 t,
\quad
x\to \lambda x, 
\quad
u_0\to \lambda^{-1} u_0,
\quad
u_{+}\to \lambda^{-p} u_{+}
\end{equation}
where $p$ is an arbitrary constant. 

The Burgers system \eqref{Burgerssys} has a potential form given by 
\begin{equation}
v_{0t} = v_{0xx} + v_{0x}^2,
\quad
v_{+t} = v_{+xx} + 2v_{+x}v_{0x}
\label{Burgerspotentialsys}
\end{equation}
from the first part of the Cole-Hopf transformation \eqref{colehopf}. 
The second part of this transformation maps this nonlinear potential system 
into a decoupled linear system of heat equations
\begin{equation}
V_{0t} = V_{0xx},
\quad
V_{+t} = V_{+xx} .
\label{linearheatsys}
\end{equation}
Correspondingly, under the same transformation, 
the hereditary recursion operator \eqref{potentialRop} 
gets mapped into the $x$-translation operator \eqref{Dxop} 
which is thereby a hereditary recursion operator for the two-component linear heat system \eqref{linearheatsys}. 

We will mention a few additional features of the integrable Burgers system \eqref{Burgerssys}. 

This system was first obtained in a classification \cite{Svi} of vector systems of Burgers form 
$\mathbf{u}_t = \mathbf{u}_{xx} + \mathbf{A}(\mathbf{u},\mathbf{u}_x)$
that possess higher symmetries.
Specifically, in \Ref{Svi}, 
see the Cole-Hopf transformation (44$'$) with parameter $c=0$; 
the corresponding nonlinear Burgers system (44) listed there is missing a factor $c$ 
in one term.

The Burgers system \eqref{Burgerssys} possesses a natural gradient flow structure, 
which is useful for analysis of solutions. 
This structure arises from the linear heat system \eqref{linearheatsys} 
into which the Burgers system is mapped by the Cole-Hopf transformation \eqref{colehopf}. 
Similarly to the ordinary heat equation, 
the two-component linear heat system \eqref{linearheatsys} 
can be expressed as a gradient flow
\begin{equation}
V_{0t} = -\frac{\delta \mk{E} }{\delta V_0},
\quad
V_{+t} = -\frac{\delta \mk{E}}{\delta V_{+}}
\end{equation}
where 
\begin{equation}\label{E}
\mk{E}= \int_{\Omega} \tfrac{1}{2}( V_{0x}^2 + V_{+x}^2 )\, dx 
\end{equation}
is a positive-definite energy functional 
on any given domain $\Omega\subseteq\Rnum$ for $x$. 
This structure is inherited by the Burgers system \eqref{Burgerssys}, 
as well as the nonlinear potential system \eqref{Burgerspotentialsys}, 
through the Cole-Hopf transformation \eqref{colehopf}
combined with the variational derivative relations
\begin{equation}\label{varderrel1}
\bmatr
\delta \mk{E}/\delta V_0 \\
\delta \mk{E}/\delta V_{+}
\ematr
= \Pop 
\bmatr
\delta \mk{E}/\delta v_0 \\
\delta \mk{E}/\delta v_{+}
\ematr,
\quad
\Pop = e^{-v_0}
\bmatr
1 & -v_{+} \\
0 & 1
\ematr
\end{equation}
and
\begin{equation}\label{varderrel2}
\bmatr
\delta \mk{E}/\delta v_0 \\
\delta \mk{E}/\delta v_{+}
\ematr 
= \Qop
\bmatr
\delta \mk{E}/\delta u_0 \\
\delta \mk{E}/\delta u_{+}
\ematr,
\quad
\Qop = 
\bmatr
-D_x & e^{v_0}v_{+} \\
0 & -D_xe^{v_0}
\ematr .
\end{equation}
Specifically, the gradient flow structure of these two nonlinear systems
is respectively given by 
\begin{equation}
\bmatr
v_0 \\
v_{+}
\ematr_t 
= -\Pop^\t \Pop 
\bmatr
\delta \mk{E}/\delta v_0 \\
\delta \mk{E}/\delta v_{+}
\ematr
\end{equation}
and
\begin{equation}
\bmatr
u_0 \\
u_{+}
\ematr_t 
= -\Kop^* \Kop
\bmatr
\delta \mk{E}/\delta u_0 \\
\delta \mk{E}/\delta u_{+}
\ematr,
\quad
\Kop= \Pop\Qop
\end{equation}
in terms of the energy functional
\begin{equation}\label{E'}
\mk{E}= \int_{\Omega} \tfrac{1}{2} e^{2v_0}\big( v_{0x}^2 + (v_{+x}+v_{+}v_{0x})^2 \big)\, dx
= \int_{\Omega} \tfrac{1}{2} \big( u_0^2e^{2D_x^{-1}u_0} + (u_{+}+u_0(D_x-u_0)^{-1}u_{+})^2 \big)\, dx 
\end{equation}
where $\Qop^\t \Qop$ is a symmetric, positive matrix,
and $\Kop^* \Kop$ is a symmetric, positive matrix operator. 
(Here, $\t$ denotes the transpose, and $*$ denotes the adjoint.)
Note $\mk{E}$ is nonlocal in terms of $(u_0,u_{+})$.

\subsection{Two-component integrable nonlinear Airy system}

All of the higher symmetries \eqref{Burgerssymms}--\eqref{Burgershierarchy} 
of the Burgers system \eqref{Burgerssys} 
correspond to integrable systems,
forming an integrable hierarchy generated by the hereditary recursion operator \eqref{Rop}. 
The first higher symmetry is given by the vector field 
$\X_{2} = f_1^{(2)}\partial_{u_0} +f_2^{(2)}\partial_{u_{+}}$
with the components 
\begin{align}
& 
\begin{aligned}
f_1^{(2)} & = (D_x +D_x u_0D_x^{-1})(u_{0xx} + 2u_0u_{0x}) 
\\\qquad
& = u_{0xxx} + 3u_0u_{0xx} + 3 u_{0x}^2 + 3u_0^2 u_{0x} ,
\end{aligned}
\\
& 
\begin{aligned}
f_2^{(2)} & = u_{+}u_{0x}-u_{0x}(D_x -u_0)^{-1}(u_{+}(u_{0xx}+u_0^2)) + (D_x -u_0)D_x(D_x -u_0)^{-1}(u_{+xx} + 2u_{+}u_{0x})
\\\qquad
& = u_{+xxx} + 3u_{+}u_{0xx} + (3 u_{0x}+ u_0^2) u_{+x} -3u_{+}u_0u_{0x} .
\end{aligned}
\end{align}
This symmetry directly corresponds to the two-component nonlinear integrable system
\begin{equation}\label{Airysys}
\begin{aligned}
& u_{0t} = u_{0xxx} + 3u_0u_{0xx} + 3 u_{0x}^2 + 3u_0^2 u_{0x} ,
\\
& u_{+t} = u_{+xxx} + 3u_{+}u_{0xx} + (3 u_{0x}+ u_0^2) u_{+x} -3u_{+}u_0u_{0x} ,
\end{aligned}
\end{equation}
which has the scaling symmetry family
\begin{equation}\label{Airyscalsymm}
t\to \lambda^3 t,
\quad
x\to \lambda x, 
\quad
u_0\to \lambda^{-1} u_0,
\quad
u_{+}\to \lambda^{-p} u_{+}
\end{equation}
where $p$ is an arbitrary constant. 

This integrable system \eqref{Airysys} describes a nonlinear version of a two-component Airy system as follows. 
Under the first part of the Cole-Hopf transformation \eqref{colehopf}, 
the system is mapped into a nonlinear potential system 
\begin{equation}\label{Airypotentialsys}
v_{0t} = v_{0xxx} + 3v_{0x}v_{0xx} + v_{0x}^3,
\quad
v_{+t} = v_{+xxx} + 3v_{0x}v_{+xx} + 3v_{+x}(v_{0xx} + v_{0x}^2) .
\end{equation}
In turn, under the second part of the Cole-Hopf transformation, 
this nonlinear potential system is mapped into a decoupled linear system of Airy equations
\begin{equation}
V_{0t} = V_{0xxx},
\quad
V_{+t} = V_{+xxx} .
\label{linearAirysys}
\end{equation}
These Airy equations do not have a gradient-flow structure 
but instead possess a bi-Hamiltonian structure. 

Recall, a linear pseudo-differential operator $\Dop$ is Hamiltonian 
with respect to a pair of variables $(w_1,w_2)$ 
iff it defines an associated Poisson bracket
\begin{equation}\label{poisson}
\{\mk{H},\mk{G} \}_{\Dop} =
\int_{\Omega} \bigg( \bmatr \delta\mk{H}/\delta w_1 \\ \delta\mk{H}/\delta w_2 \ematr^\t \Dop \bmatr \delta\mk{G}/\delta w_2 \\ \delta\mk{G}/\delta w_2 \ematr \bigg) dx
\end{equation}
obeying skew-symmetry 
$\{\mk{H},\mk{G} \}_{\Dop} + \{\mk{G},\mk{H} \}_{\Dop} =0$
and the Jacobi identity 
$\{\{\mk{H},\mk{G} \}_{\Dop},\mk{F}\}_{\Dop} + \text{cyclic } =0$, 
for all functionals $\mk{H}$, $\mk{G}$, $\mk{F}$.
The formal inverse of a Hamiltonian operator defines a symplectic operator. 
Compatibility of two Hamiltonian operators $\Dop_1$ and $\Dop_2$
is the statement that every linear combination 
$c_1\Dop_1+c_2\Dop_2$ is a Hamiltonian operator. 

The bi-Hamiltonian structure of the two-component linear Airy system \eqref{linearAirysys}
consists of 
\begin{equation}
\bmatr
V_{0} \\
V_{+} 
\ematr_t
= D_x^3
\bmatr
\delta \mk{H}/\delta V_0 \\
\delta \mk{H}/\delta V_{+}
\ematr
= -D_x 
\bmatr
\delta \mk{E}/\delta V_0 \\
\delta \mk{E}/\delta V_{+}
\ematr
\end{equation}
where $\mk{E}$ is the energy functional \eqref{E}
and $\mk{H}$ is the functional 
\begin{equation}\label{H}
\mk{H}= \int_{\Omega} \tfrac{1}{2}( V_0^2 + V_{+}^2 )\, dx
\end{equation}
given by the $L^2$ norm of $(V_0,V_{+})$
on any given domain $\Omega\subseteq\Rnum$ for $x$. 
The operators $D_x^3$ and $-D_x$ are well-known to be a compatible pair of Hamiltonian operators \cite{Olv-book}. 
By Magri's theorem \cite{Mag}, 
these Hamiltonian operators yield a hereditary recursion operator
$D_x^3(-D_x)^{-1} = -D_x^2$ for the linear Airy system \eqref{linearAirysys}. 
Note this recursion operator has the factorization $-(D_x)^2$,
where $D_x$ itself is the recursion operator arising from 
the hereditary recursion operator \eqref{Rop} of the Burgers system \eqref{Burgerssys} 
under the Cole-Hopf transformation \eqref{colehopf}. 

A similar bi-Hamiltonian formulation is inherited by 
the nonlinear Airy system \eqref{Airysys}
and its potential system \eqref{Airypotentialsys}, 
since Hamiltonian structures are preserved under invertible transformations. 
Using the variational derivative relation \eqref{varderrel1}
arising from the Cole-Hopf transformation \eqref{colehopf}, 
we get 
\begin{equation}
\bmatr
v_0 \\
v_{+}
\ematr_t 
= \wtil\Eop
\bmatr
\delta \mk{H}/\delta v_0 \\
\delta \mk{H}/\delta v_{+}
\ematr
= \wtil\Hop
\bmatr
\delta \mk{E}/\delta v_0 \\
\delta \mk{E}/\delta v_{+}
\ematr
\end{equation}
where
\begin{align}
& \wtil\Eop = -\Pop^\t D_x \Pop 
= \bmatr
-e^{-v_0}D_x e^{-v_0} & e^{-v_0}D_x e^{-v_0}v_{+} \\
-e^{-v_0}v_{+}D_x e^{-v_0} & -e^{-v_0}(v_{+}D_x e^{-v_0}v_{+} + D_x e^{-v_0})
\ematr
\label{potentialEop}
\\
& \wtil\Hop = \Pop^\t D_x^3 \Pop 
= \bmatr
e^{-v_0}D_x^3 e^{-v_0} & -e^{-v_0}D_x^3 e^{-v_0}v_{+} \\
e^{-v_0}v_{+}D_x^3 e^{-v_0} & e^{-v_0}(v_{+}D_x^3 e^{-v_0}v_{+} + D_x^3 e^{-v_0})
\ematr
\label{potentialHop}
\end{align}
are a compatible pair of Hamiltonian operators 
with respect to $(v_0,v_{+})$. 
Similarly, 
using the variational derivative relation \eqref{varderrel2}, 
we obtain 
\begin{equation}
\bmatr
u_0 \\
u_{+}
\ematr_t 
= \Eop
\bmatr
\delta \mk{H}/\delta u_0 \\
\delta \mk{H}/\delta u_{+}
\ematr
= \Hop
\bmatr
\delta \mk{E}/\delta u_0 \\
\delta \mk{E}/\delta u_{+}
\ematr
\end{equation}
where
\begin{align}
& \Eop = -\Kop^* D_x \Kop 
=\bmatr
D_x^3 & -D_x^2 e^{v_0}v_{+x} \\
-e^{v_0}v_{+x}D_x^2 & e^{v_0}(v_{+x}D_x e^{v_0}v_{+x} - D_x^3e^{v_0})
\ematr
\label{Eop}\\
& \Hop = \Kop^* D_x^3 \Kop
=\bmatr
-D_x^5 & D_x^4 e^{v_0}v_{+x} \\
e^{v_0}v_{+x}D_x^4 & e^{v_0}(D_x^5e^{v_0} -v_{+x}D_x^3 e^{v_0}v_{+x}) 
\ematr
\label{Hop}
\end{align}
are a compatible pair of Hamiltonian operators 
with respect to $(u_0,u_{+})$. 
Here $\mk{E}$ is the energy functional \eqref{E'},
while $\mk{H}$ is the positive-definite functional 
\begin{equation}\label{H'}
\mk{H}= \int_{\Omega} \tfrac{1}{2} e^{2v_0}(1 + v_{+}^2)\, dx
= \int_{\Omega} \tfrac{1}{2} \big( e^{D_x^{-1}u_0} + ((D_x-u_0)^{-1}u_{+})^2 \big)\, dx .
\end{equation}
Magri's theorem implies that the Hamiltonian operators \eqref{Eop}--\eqref{Hop} 
yield a hereditary recursion operator $\Hop\Eop^{-1}$
for the nonlinear Airy system \eqref{Airysys}. 
This recursion operator has the factorization $\Hop\Eop^{-1}=-\Rop^2$
where $\Rop$ is the hereditary recursion operator \eqref{Rop} 
for the Burgers system \eqref{Burgerssys}. 
Moreover, $\Rop$ itself is a hereditary recursion operator \eqref{Rop} 
for the nonlinear Airy system \eqref{Airysys},
since this system belongs to the hierarchy of integrable systems generated 
from the Burgers system by $\Rop$. 

The nonlinear Airy system \eqref{Airysys} is apparently new. 
It has a reduction to a one-component (scalar) system given by putting $u_+=0$. 
This reduced system first appeared in 
a classification \cite{MikShaSok} of scalar third-order evolution equations 
$u_t = A_1(x,u,u_x,u_{xx})u_{xxx} + A_0(x,u,u_x,u_{xx})$
that possess higher symmetries.
Specifically, $u_{0t} = u_{0xxx} + 3u_0u_{0xx} + 3 u_{0x}^2 + 3u_0^2 u_{0x}$ 
is equation (4.1.20) with parameters $\alpha(x)=\beta(x)=0$ in \Ref{MikShaSok}.

\subsection{Integrability structure of two-component Burgers-Airy hierarchies}

This bi-Hamiltonian structure of the nonlinear Airy system \eqref{Airysys}
is inherited by all of integrable systems that correspond to the even-order symmetries in the hierarchy \eqref{Burgerssymms}--\eqref{Burgershierarchy}. 
Likewise, all of the integrable systems that correspond to the odd-order symmetries in the hierarchy \eqref{Burgerssymms}--\eqref{Burgershierarchy}
inherit the gradient flow structure of the Burgers system \eqref{Burgerssys}. 
Both of these structures have a simple formulation in terms of $(V_0,V_{+})$:
\begin{equation}
\bmatr
V_{0} \\
V_{+} 
\ematr_t
= D_x^3
\bmatr
\delta \mk{H}^{(\frac{n}{2}-1)} /\delta V_0 \\
\delta \mk{H}^{(\frac{n}{2}-1)} /\delta V_{+}
\ematr
= -D_x 
\bmatr
\delta \mk{H}^{(\frac{n}{2})} /\delta V_0 \\
\delta \mk{H}^{(\frac{n}{2})} /\delta V_{+}
\ematr,
\quad
n=2,4,\ldots
\end{equation}
and
\begin{equation}
\bmatr
V_{0} \\
V_{+} 
\ematr_t
= -\bmatr
\delta \mk{H}^{(\frac{n-1}{2})}/\delta V_0 \\
\delta \mk{H}^{(\frac{n-1}{2})} /\delta V_{+}
\ematr,
\quad
n=3,5,\ldots
\end{equation}
where
\begin{equation}\label{Hhierarchy}
\mk{H}^{(l)}= \int_{\Omega} \tfrac{1}{2}( (D_x^l V_0)^2 + (D_x^l V_{+})^2 )\, dx,
\quad
l=0,1,2,\ldots
\end{equation}
is a higher-derivative energy functional. 
By applying the Cole-Hopf transformation \eqref{colehopf} and \eqref{colehopfX} to these structures, 
we obtain the following two results. 

\begin{thm}\label{Burgersoddhierarchy}
The hierarchy of integrable systems 
corresponding to the higher odd-order symmetries \eqref{Burgerssymms}--\eqref{Burgershierarchy}
of the Burgers system \eqref{Burgerssys}
is given by 
\begin{equation}\label{oddsys}
\bmatr
u_0 \\
u_{+}
\ematr_t
= \Rop^n
\bmatr
u_{0x} \\
u_{+x}
\ematr,
\quad
n=3,5,\ldots 
\end{equation}
(called the {\em $+\tfrac{n-1}{2}$ odd-flow})
in terms of the hereditary recursion operator \eqref{Rop}. 
Each of these integrable systems has a gradient flow structure
\begin{equation}\label{oddgradflow}
\bmatr
u_0 \\
u_{+}
\ematr_t
= -\Kop^* \Kop
\bmatr
\delta \mk{H}^{(\frac{n-1}{2})}/\delta u_0 \\
\delta \mk{H}^{(\frac{n-1}{2})}/\delta u_{+}
\ematr,
\quad
n=3,5,\ldots 
\end{equation}
where 
\begin{equation}\label{higherH'}
\mk{H}^{(l)} = \int_{\Omega} \tfrac{1}{2} \big( (D_x^l e^{v_0})^2 + (D_x^l(e^{v_0}v_{+}))^2 \big)\, dx,
\quad
l=1,2,\ldots
\end{equation}
are positive-definite energy functionals, 
and where $\Kop^* \Kop$ is a symmetric, positive matrix operator 
given by 
\begin{equation}
\Kop= \Pop\Qop
=\bmatr
-e^{-v_0}D_x & e^{-v_0}u_{+} + v_{+}(D_x+u_0) \\
0 & -(D_x+u_0) 
\ematr
\end{equation}
with $v_0=\int u_0\,dx$ and $v_{+}=\int e^{-v_0}u_{+}\,dx$ being potentials. 
\end{thm}

This gradient flow structure \eqref{oddgradflow} 
can be shown to imply 
\begin{equation}
\tfrac{d}{dt}\mk{H}^{(\frac{n-1}{2})} 
= -\int_{\Omega} \Big| 
\Kop\bmatr
\delta \mk{H}^{(\frac{n-1}{2})}/\delta u_0 \\
\delta \mk{H}^{(\frac{n-1}{2})}/\delta u_{+}
\ematr 
\Big|^2\, dx <0,
\quad
n=3,5,\ldots 
\end{equation}
(modulo boundary terms in the integral),
and so each functional $\mk{H}^{(l)}$, $l=1,2,\ldots$, 
is a positive, decreasing function of $t$.
As a consequence, 
solutions $(u_0(t,x),u_{+}(t,x))$ of each corresponding integrable system \eqref{oddsys}
having a finite norm $\mk{H}^{(l)}<\infty$ are dispersive.

\begin{thm}\label{Burgersevenhierarchy}
The hierarchy of integrable systems 
corresponding to the higher even-order symmetries \eqref{Burgerssymms}--\eqref{Burgershierarchy} 
of the Burgers system \eqref{Burgerssys}
is given by 
\begin{equation}\label{evensys}
\bmatr
u_0 \\
u_{+}
\ematr_t
= \Rop^n
\bmatr
u_{0x} \\
u_{+x}
\ematr, 
\quad
n=2,4,\ldots 
\end{equation}
(called the {\em $+\tfrac{n}{2}$ even-flow})
in terms of the hereditary recursion operator \eqref{Rop}. 
Each of these integrable systems has a bi-Hamiltonian formulation 
\begin{equation}\label{evenbiHamil}
\bmatr
u_0 \\
u_{+}
\ematr_t
= \Eop
\bmatr
\delta \mk{H}^{(\frac{n}{2}-1)}/\delta u_0 \\
\delta \mk{H}^{(\frac{n}{2}-1)}/\delta u_{+}
\ematr
= \Hop
\bmatr
\delta \mk{H}^{(\frac{n}{2})}/\delta u_0 \\
\delta \mk{H}^{(\frac{n}{2})}/\delta u_{+}
\ematr,
\quad
n=2,4,\ldots 
\end{equation}
given in terms of the energy functionals \eqref{higherH'}
and the compatible pair of Hamiltonian operators \eqref{Eop}--\eqref{Hop}. 
Each of these energy functionals is a conserved integral for all of the integrable systems \eqref{evensys}. 
\end{thm}

\subsection{Additional two-component $C$-integrable nonlinear systems and associated hierarchies}

Both the integrable Burgers system \eqref{Burgerssys}
and integrable nonlinear Air system \eqref{Airysys}
arise from applying \Thmref{structurethm} to the vector field 
$\X=u_{0x}\partial_{u_0} +u_{+x}\partial_{u_{+}}$
which generates the $x$-translation symmetry of the hereditary recursion operator \eqref{Rop}. 
This operator has additional symmetries, 
which can be used in a similar way to produce additional integrable nonlinear systems. 

The derivation is simplest if we start from the corresponding 
constant-coefficient diagonal operator \eqref{Dxop} 
obtained under the Cole-Hopf transformation \eqref{colehopf} and \eqref{colehopfX}. 
Besides possessing $x$-translation symmetry, 
generated by $\X=V_{0x}\partial_{V_0} +V_{+x}\partial_{V_{+}}$, 
this recursion operator also clearly possesses 
a rotation symmetry generated by 
\begin{equation}\label{rotation}
\X_{\text{rot.}}=V_{+}\partial_{V_0} -V_{0}\partial_{V_{+}}
\end{equation}
and a boost symmetry generated by 
\begin{equation}\label{boost}
\X_{\text{boost}} =V_{+}\partial_{V_0} +V_{0}\partial_{V_{+}}
\end{equation}
where $(V_0,V_{+})$ are the potentials \eqref{colehopf}
related to $(u_0,u_{+})$ by the Cole-Hopf transformation. 

To begin, 
we consider the boost symmetry. 
Its vector field components are given by 
\begin{equation}
H_1 =V_{+},
\quad 
H_2=V_{0}, 
\end{equation}
which we substitute into the linear evolution system \eqref{linearsys}. 
This yields the two-component system 
\begin{equation}
V_{0t} = V_{+x},
\quad
V_{+t} = V_{0x}
\end{equation}
for the potentials. 
Combining these two equations, 
we obtain two decoupled wave equations
\begin{equation}\label{linearwavesys}
V_{0tt} = V_{0xx},
\quad
V_{+tt} = V_{+xx} .
\end{equation}
Clearly, this two-component linear wave system possesses the same 
constant-coefficient diagonal operator \eqref{Dxop} 
as the linear heat system  \eqref{linearheatsys} and the linear Airy system \eqref{linearAirysys}. 
Going back to the intermediate potentials $(v_0,v_{+})$ in the Cole-Hopf transformation \eqref{colehopf}, 
we obtain the wave system 
\begin{equation}\label{2ndordpotentialwavesys}
\begin{aligned}
v_{0tt} & = v_{0xx} +(1-v_{+}^2)v_{0x}^2 -2v_{+}v_{0x}v_{+x} -v_{+x}^2 ,
\\
v_{+tt} & = v_{+xx} +2v_{+}v_{+x}^2 +4v_{+}^2v_{0x}v_{+x} +2(1-v_{+x}^2)v_{+}v_{0x}^2 ,
\end{aligned}
\end{equation}
which possesses the recursion operator \eqref{potentialRop}.
Note the nonlinearities have a semilinear form. 

This semilinear wave system \eqref{2ndordpotentialwavesys} has the following additional integrability properties. 
Firstly, it possesses a hierarchy of higher-order symmetries 
generated from the recursion operator \eqref{potentialRop} 
applied to the $x$-translation symmetry vector field
$\X=v_{0x}\partial_{v_0} +v_{+x}\partial_{v_{+}}$. 
Secondly, when the system is expressed in its equivalent first-order form 
\begin{equation}\label{potentialwavesys}
v_{0t} = v_{+}v_{0x} +v_{+x}, 
\quad
v_{+t}  = (1-v_{+x}^2)v_{0x} -v_{+}v_{+x} , 
\end{equation}
it has a bi-Hamiltonian structure given by 
\begin{equation}
\bmatr
v_0 \\
v_{+}
\ematr_t 
= \wtil\Dop_1
\bmatr
\delta \mk{H}/\delta v_0 \\
\delta \mk{H}/\delta v_{+}
\ematr
= \wtil\Dop_2
\bmatr
\delta \mk{E}/\delta v_0 \\
\delta \mk{E}/\delta v_{+}
\ematr
\end{equation}
where
\begin{align}
& \wtil\Dop_1 = -\Pop^\t \bmatr 0 &  D_x \\ D_x & 0 \ematr \Pop 
= \bmatr
0 & -e^{-v_0} D_x e^{-v_0} \\
-e^{-v_0} D_x e^{-v_0} & e^{-v_0}(D_x e^{-v_0}v_{+} + v_{+}D_x e^{-v_0})
\ematr
\\
& \wtil\Dop_2 = -\Pop^\t \bmatr 0 &  D_x^{-1} \\ D_x^{-1} & 0 \ematr \Pop 
= \bmatr
0 & e^{-v_0} D_x^{-1} e^{-v_0} \\
e^{-v_0} D_x^{-1} e^{-v_0} & -e^{-v_0}(D_x^{-1} e^{-v_0}v_{+} + v_{+}D_x^{-1} e^{-v_0})
\ematr
\end{align}
are a compatible pair of Hamiltonian operators with respect to $(v_0,v_{+})$,
and where $\mk{E}$ is the energy functional \eqref{E'} 
and $\mk{H}$ is the positive-definite functional \eqref{H'}, respectively. 
Note $\Pop$ is the operator shown in equation \eqref{varderrel1},
which relates variational derivatives with respect to $(v_0,v_{+})$ and $(V_0,V_{+})$. 

In terms of the Cartan matrix components $(u_0,u_{+})$, 
the semilinear wave system \eqref{2ndordpotentialwavesys} 
as well as the equivalent first-order system \eqref{potentialwavesys} 
are nonlocal. 
Explicitly, 
we have 
\begin{align}
& u_{0t} = v_{+}u_{0x} +e^{-v_0}u_{+x}, 
\label{wavesys1}
\\
& u_{+t}  = (1-v_{+x}^2)e^{v_0}u_{0x} -v_{+}u_{+x} , 
\label{wavesys2}
\end{align}
and 
\begin{equation}\label{2ndordwavesys}
\begin{aligned}
u_{0tt} & = u_{0xx} +2(1-v_{+}^2)u_0u_{0x} -2e^{-v_0}v_{+}(u_0 u_{+})_x  -2e^{-2v_0}u_{+}u_{+x} , 
\\
u_{+tt} & = u_{+xx} +2e^{-v_0}v_{+}u_{+}u_{+x} +2(v_{+}^2-1)( (u_0u_{+})_x + e^{v_0}v_{+}u_0u_{0x}) .
\end{aligned}
\end{equation}
The corresponding bi-Hamiltonian structure of the first-order system \eqref{wavesys1}--\eqref{wavesys2}
is given by 
\begin{equation}
\bmatr
u_0 \\
u_{+}
\ematr_t 
= \Dop_1
\bmatr
\delta \mk{H}/\delta u_0 \\
\delta \mk{H}/\delta u_{+}
\ematr
= \Dop_2
\bmatr
\delta \mk{E}/\delta u_0 \\
\delta \mk{E}/\delta u_{+}
\ematr
\end{equation}
with 
\begin{align}
&\begin{aligned} 
\Dop_1 & = \Qop^* \wtil\Dop_1 \Qop = -\Kop^*  \bmatr 0 &  D_x \\ D_x & 0 \ematr \Kop 
\\& 
= \bmatr 
0 & -(D_x e^{-v_0})^2 D_x e^{v_0} \\
e^{v_0} (D_x e^{-v_0})^2 D_x & 
\begin{aligned}
& v_{+} D_x (D_x +u_0) -(D_x -u_0)D_x v_{+} 
\\&\quad - (D_x-u_0)(v_{+}D_x +D_x v_{+})(D_x+u_0) 
\end{aligned} 
\ematr ,
\end{aligned} 
\label{Dop1}
\\
& \begin{aligned} 
\Dop_2  & = \Qop^* \wtil\Dop_2 \Qop = -\Kop^*  \bmatr 0 &  D_x^{-1} \\ D_x^{-1} & 0 \ematr \Kop 
\\& 
= \bmatr
0 & D_x e^{-v_0}D_x^{-1}(D_x +u_0) \\
-(D_x -u_0)D_x^{-1} e^{-v_0}D_x & 
\begin{aligned}
& (D_x -u_0)D_x^{-1} v_{+} - v_{+} D_x^{-1} (D_x +u_0) 
\\&\quad + (D_x-u_0)(v_{+}D_x^{-1} +D_x^{-1} v_{+})(D_x+u_0) 
\end{aligned} 
\ematr ,
\end{aligned} 
\label{Dop2}
\end{align}
which are compatible Hamiltonian operators. 

Next, we consider the rotation symmetry, 
whose vector field components are given by 
\begin{equation}
H_1 =V_{+},
\quad 
H_2=-V_{0}
\end{equation}
We substitute these components into the linear evolution system \eqref{linearsys}
which yields the two-component system 
\begin{equation}
V_{0t} = V_{+x},
\quad
V_{+t} = -V_{0x} .
\end{equation}
for the potentials. 
A more interesting system is obtained if we apply the recursion operator \eqref{Dxop},
which produces a linear Schr\"odinger system 
\begin{equation}\label{linearSchrsys}
V_{0t} = V_{+xx},
\quad
V_{+t} = -V_{0xx} .
\end{equation}
In particular, 
$V=V_0+iV_{+}$ satisfies the linear Schr\"odinger equation $i V_t = V_{xx}$. 

The linear Schr\"odinger system \eqref{linearSchrsys} possesses 
the same recursion operator \eqref{Dxop} 
as the linear wave system  \eqref{linearwavesys}. 
Going back to the intermediate potentials $(v_0,v_{+})$ in the Cole-Hopf transformation \eqref{colehopf}, 
we obtain the Schr\"odinger system 
\begin{equation}\label{potentialSchrsys}
\bmatr 
v_0 \\ v_{+} 
\ematr_t
= \bmatr
1 & -v_{+} \\
-v_{+} & 1+v_{+}^2
\ematr
\left(
\bmatr
v_{+xx} \\ -v_{0xx} 
\ematr
+ 
v_{0x}
\bmatr
2 v_{+x} \\ -v_{0x}
\ematr
\right)
\end{equation}
which possesses the recursion operator \eqref{potentialRop}.
Note the nonlinearities have a quasilinear form. 
We can express this quasilinear Schr\"odinger system \eqref{potentialSchrsys}
in a complex-variable form by introducing the following matrices:
\begin{equation}
J = \bmatr 0 & -1 \\ 1 & 0 \ematr,
\quad
J^2 = -I
\end{equation}
which represents multiplication by $i$, 
where $I$ denotes the identity matrix;
\begin{equation}
\wtil K(v_{+}) = \bmatr
1+v_{+}^2 & v_{+} \\
v_{+} & 1
\ematr ,
\quad
\wtil K^\t = \wtil K,
\quad
\det(\wtil K)=1
\end{equation}
which is positive, symmetric, and unimodular;
and
\begin{equation}
\wtil C(v_{+}) = \bmatr
1+v_{+}^2 & 2v_{+} \\
v_{+} & 2
\ematr
\end{equation}
which is positive. 
Then we have
\begin{equation}\label{potentialSchrsys-complexvar}
J \bmatr 
v_0 \\ v_{+} 
\ematr_t
= 
\wtil K(v_{+})
\bmatr
v_0 \\ v_+
\ematr_{xx} 
+ 
v_{0x}\wtil C(v_{+})
\bmatr
v_0\\ v_+
\ematr_{x} 
\end{equation}
where $(v_0,v_{+})$ can be viewed as a complex variable $v=v_0+iv_{+}$. 

The quasilinear Schr\"odinger system \eqref{potentialSchrsys} 
has the following additional integrability properties. 
Firstly, it possesses the same hierarchy of higher-order symmetries 
as the semilinear wave system \eqref{potentialwavesys}. 
Secondly, it has a bi-Hamiltonian structure given by 
\begin{equation}
\bmatr
v_0 \\
v_{+}
\ematr_t 
= \wtil\Eop
\bmatr
\delta \mk{J}/\delta v_0 \\
\delta \mk{J}/\delta v_{+}
\ematr
= \wtil\Jop
\bmatr
\delta \mk{E}/\delta v_0 \\
\delta \mk{E}/\delta v_{+}
\ematr
\end{equation}
with a compatible pair of Hamiltonian operators with respect to $(v_0,v_{+})$,
where $\wtil\Eop$ is the Hamiltonian operator \eqref{potentialEop}, 
$\wtil\Jop$ is a Hamiltonian operator given by 
\begin{equation}
\wtil\Jop = \Pop^\t J \Pop 
= \bmatr 
0 & -e^{2v_0} \\
e^{2v_0} & 0
\ematr 
= e^{2v_0} J, 
\end{equation}
and where $\mk{E}$ is the energy functional \eqref{E'} 
and $\mk{J}$ is the momentum functional 
\begin{equation}\label{J'}
\mk{J}= \int_{\Omega} e^{2v_0} v_{+}v_{0x}\, dx
= \int_{\Omega} e^{D_x^{-1}u_0}u_0(D_x-u_0)u_{+} \, dx .
\end{equation}

When the quasilinear Schr\"odinger system \eqref{potentialSchrsys} is expressed 
in terms of the Cartan matrix components $(u_0,u_{+})$, 
it becomes nonlocal
\begin{equation}\label{Schrsys}
\bmatr 
u_0 \\ u_{+} 
\ematr_t
= \bmatr
e^{-v_0} & -v_{+} \\
-v_{+} & e^{v_0}(1+v_{+}^2)
\ematr
\left(
\bmatr
u_{+xx} \\ -u_{0xx} 
\ematr
+ 
2u_{0x}
\bmatr
u_{+x} \\ -u_{0x}
\ematr
\right)
\end{equation}
which has the equivalent form
\begin{equation}\label{Schrsys-complexvar}
J \bmatr 
u_0 \\ u_{+} 
\ematr_t
= 
K(v_{+})
\left(
\bmatr
u_0 \\ u_+
\ematr_{xx} 
+ 
2u_{0x}
\bmatr
u_0\\ u_+
\ematr_{x} 
\right)
\end{equation}
with 
\begin{equation}
K(v_{+}) = \bmatr
e^{v_0}(1+v_{+}^2) & v_{+} \\
v_{+} & e^{-v_0}
\ematr ,
\quad
K^\t = K,
\quad
\det(K)=1
\end{equation}
being a positive, symmetric, and unimodular matrix, 
where $(u_0,u_{+})$ can be viewed as a complex variable $u=u_0+iu_{+}$. 
The bi-Hamiltonian structure of this system is given by 
\begin{equation}
\bmatr
u_0 \\
u_{+}
\ematr_t 
= \Eop
\bmatr
\delta \mk{J}/\delta u_0 \\
\delta \mk{J}/\delta u_{+}
\ematr
= \Jop
\bmatr
\delta \mk{E}/\delta u_0 \\
\delta \mk{E}/\delta u_{+}
\ematr
\end{equation}
where $\Eop$ is the Hamiltonian operator \eqref{Eop},
and $\Jop$ is the Hamiltonian operator
\begin{equation}\label{Jop}
\Jop = \Qop^* \wtil\Jop \Qop = \Kop^* J\Kop 
= \bmatr 
0 & D_x e^{v_0}(D_x+u_0) \\
-(D_x-u_0) e^{-v_0}D_x & v_{+}(D_x+u_0) + (D_x-u_0)v_{+} .
\ematr 
\end{equation}

Both the quasilinear Schr\"odinger system \eqref{Schrsys} 
and the semilinear wave system \eqref{wavesys1}--\eqref{wavesys2} 
possess the hierarchy of higher symmetries \eqref{Burgerssymms}--\eqref{Burgershierarchy} admitted by the Burgers system \eqref{Burgerssys}. 
Additionally, 
each of these systems \eqref{Schrsys} and \eqref{wavesys1}--\eqref{wavesys2} 
possesses a related hierarchy of higher symmetries that are generated by  
the hereditary recursion operator applied to the rotation and boost vector fields
\eqref{rotation}--\eqref{boost}
expressed in terms of $(u_0,u_{+})$ through the Cole-Hopf transformation \eqref{colehopfX}. 
The resulting higher-order symmetries correspond to 
two hierarchies of integrable systems, 
each of which is a higher-derivative version of 
either the quasilinear Schr\"odinger system \eqref{Schrsys} 
or the semilinear wave system \eqref{wavesys1}--\eqref{wavesys2},
respectively. 

This leads to the following result. 

\begin{thm}\label{additionalhierarchies}
(i)
The semilinear wave system \eqref{wavesys1}--\eqref{wavesys2} 
and the quasilinear Schr\"odinger system \eqref{Schrsys}
belong to respective hierarchies of two-component integrable systems 
\begin{equation}
\bmatr
u_0 \\
u_{+}
\ematr_t
= \Rop^{2n}
\bmatr f_1\\ f_2 \ematr, 
\quad
n=0,1,2,\ldots 
\end{equation}
in terms of the hereditary recursion operator \eqref{Rop},
where $(f_1,f_2)$ is given by the right hand sides of 
the respective coupled equations \eqref{wavesys1}--\eqref{wavesys2} and \eqref{Schrsys}. 
\newline
(ii) 
These two hierarchies have a bi-Hamiltonian formulation. 
In the case of the semilinear wave system hierarchy, 
this structure is given by 
\begin{equation}\label{wavesyshierarchy}
\bmatr
u_0 \\
u_{+}
\ematr_t
= \Dop_1
\bmatr
\delta \mk{H}^{(n)}/\delta u_0 \\
\delta \mk{H}^{(n)}/\delta u_{+}
\ematr
= \Dop_2
\bmatr
\delta \mk{H}^{(n+1)}/\delta u_0 \\
\delta \mk{H}^{(n+1)}/\delta u_{+}
\ematr,
\quad
n=0,1,2,\ldots 
\end{equation}
using the higher-derivative energy functionals \eqref{higherH'}
and the compatible pair of Hamiltonian operators \eqref{Dop1}--\eqref{Dop2}. 
Each of these energy functionals is a conserved integral for all of the integrable systems \eqref{wavesyshierarchy}. 
In the case of the quasilinear Schr\"odinger system hierarchy, 
its bi-Hamiltonian structure is given by 
\begin{equation}\label{Schrsyshierarchy}
\bmatr
u_0 \\
u_{+}
\ematr_t
= \Eop
\bmatr
\delta \mk{J^{(n)}}/\delta u_0 \\
\delta \mk{J^{(n)}}/\delta u_{+}
\ematr
= \Jop
\bmatr
\delta \mk{H^{(n+1)}}/\delta u_0 \\
\delta \mk{H^{(n+1)}}/\delta u_{+}
\ematr,
\quad
n=0,1,2,\ldots 
\end{equation}
using the compatible pair of Hamiltonian operators \eqref{Eop} and \eqref{Jop},
and the higher-derivative energy functionals \eqref{higherH'}
along with the higher-derivative momentum functionals 
\begin{equation}\label{higherJ'}
\mk{J}^{(l)} = \int_{\Omega} \tfrac{1}{2} (D_x^{l+1} e^{v_0})D_x^l(e^{v_0}v_{+})\, dx,
\quad
l=0,1,2,\ldots .
\end{equation}
Each of these functionals is a conserved integral for all of the integrable systems \eqref{Schrsyshierarchy}. 
\end{thm}

\section{A Hasimoto transformation for null curves in $\Mink{2}$}
\label{sec:gauge}

We will now discuss the geometrical meaning of the gauge choice \eqref{gaugecond}
imposed on the Cartan matrix \eqref{nulltangentU} of a general null-tangent frame \eqref{nulltangentframe}
which we have used in obtaining \Thmref{structurethm}
for null curve flows in $\Mink{2}$. 

Consider the action of an $x$-dependent $SO(2,1)$ transformation group 
on the frame vectors 
\begin{equation}
\wtil\E = 
\bmatr
\tilde\e_{+} \\
\tilde\e_{-} \\
\tilde\e_{\perp}
\ematr
\end{equation}
comprising a general null-tangent frame \eqref{nulltangentframe}
for null curves with an arbitrary smooth parameterization $x$. 
For one-dimensional subgroups with the action 
\begin{equation}\label{Etransformation}
\wtil\E \to \exp(\epsilon(x)S)\wtil\E ,
\end{equation}
the generator is given by 
\begin{equation}
S = \bmatr 
s_0 & 0 & s_{+} \\ 
0 & -s_0 & s_{-} \\
s_{-} & s_{+} & 0
\ematr
\end{equation}
which is a matrix representation of the Lie algebra $\mk{so}(2,1)$
with respect to a null-tangent frame. 
Here $\epsilon(x)$ is an arbitrary function of the parameter $x$ 
along the null curve. 
We can decompose $S=S_{\perp}+S_{-}+S_{+}$ into matrices 
\begin{equation}
S_{\perp} = \bmatr 
s_0 & 0 & 0 \\ 
0 & -s_0 & 0 \\
0 & 0 & 0
\ematr ,
\quad
S_{-} = 
\bmatr
0 & 0 & s_{+} \\ 
0 & 0 & 0 \\
0 & s_{+} & 0
\ematr,
\quad
S_{+} = 
\bmatr
0 & 0 & 0 \\ 
0 & 0 & s_{-} \\
s_{-} & 0 & 0
\ematr ,
\end{equation}
which have the respective properties 
\begin{equation}
S_{\perp}\bmatr
0 \\
0 \\
\tilde\e_{\perp}
\ematr
=0,
\quad
S_{-}\bmatr
0 \\
\tilde\e_{-} \\
0 \\
\ematr
=0,
\quad
S_{+}\bmatr
\tilde\e_{+} \\
0 \\
0 \\
\ematr
=0 .
\end{equation}
These matrices generate a corresponding decomposition of the Lie algebra $\mk{so}(2,1)$ 
into a direct sum of one-dimensional subalgebras
\begin{equation}\label{liealgdecomp}
\mk{so}(2,1) = \mk{so}(2,1)_{\perp}\oplus \mk{so}(2,1)_{-}\oplus \mk{so}(2,1)_{+}
\end{equation}
with the Lie bracket structure
\begin{equation}
[\mk{so}(2,1)_{\perp},\mk{so}(2,1)_{\pm}] \subseteq \mk{so}(2,1)_{\pm},
\quad
[\mk{so}(2,1)_{+},\mk{so}(2,1)_{-}] \subseteq \mk{so}(2,1)_{\perp},
\end{equation}
where 
\begin{equation}
\mk{so}(2,1)_{\perp} =\spn(S_{\perp}),
\quad
\mk{so}(2,1)_{-} =\spn(S_{-}),
\quad
\mk{so}(2,1)_{+} =\spn(S_{+})
\end{equation}
are generators of one-dimensional transformation groups that are stabilizers of $\tilde\e_{\perp}$, $\tilde\e_{-}$, $\tilde\e_{+}$, respectively. 

Then we see that the form of the Cartan matrix \eqref{nulltangentUWgauge}
in the gauge \eqref{gaugecond} 
is characterized by the property 
\begin{equation}\label{isotropyprop}
\wtil\U \in \mk{so}(2,1)_{\perp}\oplus\mk{so}(2,1)_{-} .
\end{equation}
This can be stated more directly as 
\begin{equation}\label{gauge}
\wtil\U = \wtil\U_{\perp} + \wtil\U_{-} + \wtil\U_{+}, 
\quad
\wtil\U_{+} =0
\end{equation}
where $\wtil\U_{\perp}$ and $\wtil\U_{\pm}$ are the components of the Cartan matrix 
relative to the decomposition \eqref{liealgdecomp}. 
Thus, 
$\wtil\U$ has no components belonging to the subalgebra of $\mk{so}(2,1)$ 
that leaves invariant the tangent vector $\tilde\e_{+}$ in the null frame
(namely, $\mk{so}(2,1)_{+}$ is the isotropy subalgebra for $\tilde\e_{+}$). 
We will therefore call this a \emph{minimal-isotropy} gauge,
and the resulting null-tangent frame will be referred to as a minimal-isotropy null frame. 

The geometrical meaning of this gauge \eqref{gauge}
can be understood by looking at how the frame vectors are transported by 
the resulting Cartan matrix \eqref{nulltangentUWgauge} along the null curve. 
We see that the spacelike normal vector $\tilde\e_{\perp}$ 
undergoes a boost in the direction of the null vector $\tilde\e_{-}$,
and this null vector gets scaled, 
while the null tangent vector $\tilde\e_{+}$ is scaled and also boosted 
in the normal direction given by $\tilde\e_{\perp}$. 

\Thmref{structurethm} shows that we can obtain a minimal-isotropy null frame
by applying a suitable little-group transformation \eqref{gaugegroup}--\eqref{SonE}
to a Frenet null frame \eqref{Frenetnullframe}. 
The transformed frame vectors \eqref{nulltangentframe} are explicitly given by 
\begin{equation}\label{nullparallelframe}
\tilde\e_{+} = \T, 
\quad
\tilde\e_{-} = \B +\epsilon \N +\tfrac{1}{2} \epsilon^2 \T,
\quad
\tilde\e_{\perp} = \N + \epsilon \T
\end{equation}
in terms of the group parameter $\epsilon(x)$ 
which is given by any solution of the Riccati equation \eqref{gaugerel1}. 
This group parameter $\epsilon(x)$ will contain an arbitrary integration constant
which represents a residual gauge freedom in the little-group transformation
\eqref{gaugegroup}--\eqref{SonE} as follows. 
By combining the Riccati equation \eqref{gaugerel1} 
and the expression \eqref{u0u+} relating $u_0$ to $\epsilon(x)$, 
we see that $u_0$ satisfies the similar Riccati equation
\begin{equation}\label{u0Riccati}
(u_0/\kappa)_x + u_0^2/(2\kappa) +\tau =0
\end{equation}
in terms of the Frenet curvature $\kappa$ and torsion $\tau$. 
Every Riccati equation possesses a one-parameter transformation group 
on its solutions. 
When this transformation group is applied to equation \eqref{u0Riccati}, 
we obtain 
\begin{equation}\label{u0Riccatigroup}
u_0\to u_0 + 2\varepsilon v_{+}/(1+\varepsilon v_{+})
\end{equation}
where $\varepsilon$ is an arbitrary constant parameter,
and where $v_{+} = \int e^{-v_0}\kappa\,dx$ and $v_0 = \int u_0\,dx$ 
are the potentials \eqref{u+potential} and \eqref{u0potential}
which appear in the Cole-Hopf transformation. 
Note $\varepsilon=0$ corresponds to the identity transformation on $u_0$. 

We will now show that the one-parameter transformation group \eqref{u0Riccatigroup}
geometrically represents a residual little group of rigid gauge transformations 
under which the gauge conditions \eqref{gaugecond} and \eqref{flowcond}
are preserved. 
Consider the little group of gauge transformations \eqref{gaugegroup}
and let $\epsilon=\tilde\epsilon(x,t)$ be an arbitrary gauge parameter. 
The gauge conditions \eqref{gaugecond} and \eqref{flowcond} 
imply that the Cartan matrix \eqref{nulltangentU} 
and the Cartan flow matrix \eqref{nulltangentW} 
have the respective forms
\begin{equation}\label{nulltangentUWgauge}
\wtil\U=
\bmatr 
u_0 & 0 & u_{+} \\ 
0 & -u_0 & 0 \\ 
0 & u_{+} & 0 
\ematr,
\quad
\wtil\W=
\bmatr 
\omega_0 & 0 & \omega_{+} \\ 
0 & -\omega_0 & 0 \\ 
0 & \omega_{+} & 0 
\ematr .
\end{equation}
Under the little gauge group, 
these matrices transform as 
\begin{equation}\label{gaugedUW}
\wtil\U \to (\S(\tilde\epsilon)_x+\S(\tilde\epsilon)\wtil\U)\S(\tilde\epsilon)^{-1}
= \bmatr 
\tilde u_0 & 0 & \tilde u_{+} \\ 
0 & -\tilde u_0 & \tilde u_{-} \\ 
\tilde u_{-} & \tilde u_{+} & 0
\ematr
\end{equation}
and
\begin{equation}
\wtil\W \to (\S(\tilde\epsilon)_t+\S(\tilde\epsilon)\wtil\W)\S(\tilde\epsilon)^{-1}= 
\bmatr 
\tilde\omega_0 & 0 & \tilde\omega_{+} \\ 
0 & -\tilde\omega_0 & \tilde\omega_{-} \\ 
\tilde\omega_{-} & \tilde\omega_{+} & 0 
\ematr
\end{equation}
where
\begin{equation}\label{u+u0littlegroup}
\tilde u_{+}= u_{+},
\quad
\tilde u_0= u_0 -\tilde\epsilon u_{+},
\quad
\tilde u_{-} = \tilde\epsilon_x +\tilde\epsilon u_0 -\tfrac{1}{2}\tilde\epsilon^2 u_{+}
\end{equation}
and
\begin{equation}\label{w+w0littlegroup}
\tilde\omega_{+}=\omega_{+},
\quad
\tilde\omega_0=\omega_0 -\tilde\epsilon\omega_{+},
\quad
\tilde\omega_{-} = \tilde\epsilon_t +\tilde\epsilon\omega_0 -\tfrac{1}{2}\tilde\epsilon^2\omega_{+} .
\end{equation}
Preservation of the form of $\wtil\U$ and $\wtil\W$ 
requires the conditions $\tilde u_{-}=0$ and $\tilde w_{-}=0$, 
which yields 
\begin{equation}
\tilde\epsilon_x +\tilde\epsilon u_0 -\tfrac{1}{2}\tilde\epsilon^2 u_{+}=0,
\quad
\tilde\epsilon_t +\tilde\epsilon\omega_0 -\tfrac{1}{2}\tilde\epsilon^2\omega_{+} =0 .
\end{equation}
It is useful to express $(u_0,u_{+})$ and $(\omega_0,\omega_{+})$ 
in terms of the potentials \eqref{u0potential} and \eqref{u+potential}.
This gives 
\begin{equation}
\tilde\epsilon_x +\tilde\epsilon v_{0x} -\tfrac{1}{2}\tilde\epsilon^2 e^{v_0}v_{+x}=0,
\quad
\tilde\epsilon_t +\tilde\epsilon v_{0t} -\tfrac{1}{2}\tilde\epsilon^2 e^{v_0}v_{+t} =0
\end{equation}
which are a compatible pair of Bernoulli equations for $\tilde\epsilon(x,t)$. 
The general solution is given by 
\begin{equation}
\tilde\epsilon = -2\varepsilon e^{-v_0}/(1+\varepsilon v_{+})
\end{equation}
where $\varepsilon$ is an arbitrary constant. 
Then the transformations \eqref{u+u0littlegroup} and \eqref{w+w0littlegroup} 
preserve the gauge conditions 
and leave invariant $u_{+}$ and $\omega_{+}$,
while their action on $u_0$ and $\omega_0$ is given by 
\begin{equation}\label{u0v0littlegroup}
\tilde u_0= u_0 +2\varepsilon v_{+x}/(1+\varepsilon v_{+}), 
\quad
\tilde\omega_0=\omega_0 +2\varepsilon v_{+t}/(1+\varepsilon v_{+}) .
\end{equation}
This is the same as the transformation group \eqref{u0Riccatigroup}
arising from the Riccati equation \eqref{u0Riccati}. 
The resulting little gauge group has the matrix representation 
\begin{equation}
\S(\tilde\epsilon) 
= \bmatr 
1 & 0 & 0 \\ 
2\varepsilon^2e^{-2v_0}/(1+\varepsilon v_{+})^2 & 1 & -2\varepsilon e^{-v_0}/(1+\varepsilon v_{+}) \\
-2\varepsilon e^{-v_0}/(1+\varepsilon v_{+}) & 0 & 1
\ematr .
\end{equation}

Finally, we remark that this gauge group is rigid in the sense that 
freely specifying the value of the gauge parameter $\tilde\epsilon$ at any point $x=x_0$ 
fixes the constant $\varepsilon$ 
whereby $\S(\tilde\epsilon)$ is then uniquely determined at all points $x$.

\subsection{Relationship with Euclidean parallel frames}

In Euclidean space $\Rnum^3$,
for curves $\pos(x)$ with an arbitrary smooth parameterization $x$, 
the closest analog of a minimal-isotropy frame is a parallel frame. 
The construction of a parallel frame starts from a Frenet frame 
given in terms of the tangent vector $\T=\pos_x$ and the arclength $s=\int |\T|\, dx$ 
by $(\what\T,\what\N,\what\B)$ 
where $\what\T = |\T|^{-1}\T$ is the unit tangent vector, 
$\what\N=|\T_x|^{-1}\T_x$ is the unit normal vector, 
and $\what\B = \what\T\times\what\N$ is the unit bi-normal vector. 
Then a $SO(2)$ gauge transformation is applied to the vectors $\what\N$ and $\what\B$,
such that the Cartan matrix of the transformed frame 
has the following geometric and algebraic characterizations.
Geometrically, 
the transport of the two vectors orthogonal to $\T$ in the parallel frame along the curve $\pos(x)$ 
is given by a rotation of each of these vectors in the plane containing 
the vector itself and the tangent vector $\T$. 
This property is analogous to the boosting of both the normal vector $\tilde\e_{\perp}$ and the tangent vector $\tilde\e_{+}$ 
when a minimal-isotropy frame is transported along a null curve. 
Algebraically, 
the Cartan matrix of a parallel frame belongs to the perp space of the isotropy subalgebra
$\mk{so}(2)\subset \mk{so}(3)$ of the $SO(2)$ group that preserves the tangent vector,
where $\mk{so}(3)$ is the Lie algebra of the $SO(3)$ rotation group acting on the frame. 
This property is a direct analog of the algebraic property \eqref{isotropyprop}
that characterizes a minimal-isotropy gauge. 

The $SO(2)$ gauge transformation relating a parallel frame to a Frenet frame 
corresponds to the well-known Hasimoto transformation
$u=u_1+iu_2 =\kappa \exp(i\int\tau\,dx)$ 
where $(\kappa,\tau)$ are the curvature and torsion invariants of the curve,
and $(u_1,u_2)$ are components of the Cartan matrix of the parallel frame. 
This transformation is important for mapping 
the evolution equations on $(\kappa,\tau)$ given by 
the vortex filament curve flow equation 
$\pos_t = \kappa \B$ 
and its axial generalization 
$\pos_t = \tfrac{1}{2}\kappa^2\T + \kappa_x\N +\kappa\tau\B$
into integrable systems consisting of, respectively, 
the focusing NLS equation 
$-iu_t = u_{xx} + \tfrac{1}{2}|u|^2u$ 
and the focusing complex mKdV equation 
$u_t = u_{xxx} + \tfrac{3}{2}|u|^2u_x$. 
Both of these integrable systems are invariant 
under rigid $U(1)$ phase rotations $u\to e^{i\phi}u$, 
involving an arbitrary constant phase angle $\phi$. 
This invariance geometrically corresponds to the residual gauge freedom 
inherent in the form of a parallel frame. 
In particular, for a given arclength-parameterized curve in Euclidean space, 
any two parallel frames are related by a rigid $SO(2)$ rotation 
which induces a $U(1)$ phase rotation on the Cartan matrix components $u=u_1+iu_2$. 
Hence, $u$ represents a $U(1)$-covariant of the curve, 
in contrast to the invariants $(\kappa,\tau)$ 
which are determined uniquely by the curve. 

A similar situation arises in Minkowski space $\Mink{2}$ 
when non-null curves are considered \cite{AncAlk}. 
In the case of timelike curves, with a proper-time parameterization, 
the transformation from a Frenet frame to a parallel frame 
corresponds to the same Hasimoto transformation 
$u=u_1+iu_2 =\kappa \exp(i\int\tau\,dx)$ 
as in Euclidean space, 
where rigid $U(1)$ phase rotations $u\to e^{i\phi}u$
represent the residual $SO(2)$ gauge freedom in the form of a parallel frame. 
In the case of spacelike curves, with a proper-length parameterization, 
the analogous Hasimoto transformation is a hyperbolic generalization 
$u=u_1+ju_2 =\kappa \exp(j\int\tau\,dx)$ 
based on the split-complex numbers \cite{splitcomplex}
defined by $j^2 = 1$ and $\bar{j} =-j$.
Any two parallel frames are related by a $SO(1,1)$ boost 
(namely, a hyperbolic rotation), 
which induces a $SO(1,1)$ hyperbolic phase rotation 
$u\rightarrow u\exp(j\phi)$ on the Cartan matrix components $u=u_1+ju_2$. 
This split-complex variable $u$ geometrically represents a $SO(1,1)$-covariant of the curve.

\subsection{Geometrical properties}

For parameterized null curves in $\Mink{2}$, 
the little-group gauge transformation shown in \Thmref{structurethm}
can be interpreted in a similar way as a Hasimoto transformation. 
In particular, under this gauge transformation, 
a Frenet null frame \eqref{Frenetnullframe} is transformed to a minimal-isotropy null frame \eqref{nullparallelframe} 
which is unique up to a residual gauge transformation \eqref{u0Riccatigroup}
given by a rigid representation of the little group. 
This residual gauge transformation can be written explicitly as 
a Riccati transformation 
\begin{equation}\label{Ucovariance}
\bmatr
u_{+} \\
u_0 
\ematr
\to 
\bmatr
u_{+} \\
u_0+2\varepsilon\int (u_{+}/u_0)dx/\big(1+\varepsilon\int (u_{+}/u_0)dx\big)
\ematr
\end{equation}
where $\varepsilon$ is an arbitrary constant parameter. 
The relationship \eqref{u0u+}
expressing this pair of Cartan matrix components $(u_{+},u_0)$ 
in terms of the Frenet curvature and torsion $(\kappa,\tau)$ of the parameterized null curve
constitutes a Hasimoto transformation,
and geometrically, 
$u_0$ represents a little-group covariant of the null curve
while $u_{+}$ represents an invariant,
relative to the arbitrary parameterization. 
When this Hasimoto transformation is inverted, 
the expressions
\begin{equation}\label{curvtorssys}
\kappa = u_{+},
\quad
\tau = -(u_0/u_{+})_x - \tfrac{1}{2} (u_0^2/u_{+})
\end{equation}
for Frenet curvature and torsion are invariant under the Riccati transformation \eqref{Ucovariance}. 
This is analogous to what happens in the case of non-null curves. 

One difference between this transformation \eqref{Ucovariance}
and the residual gauge freedom in Hasimoto transformations in the case of non-null curves
is that it is nonlinear and nonlocal in the variables $(u_{+},u_0)$. 
As a consequence, when the infinitesimal generator is written as a vector field
$\X = 2\int (u_{+}/u_0)dx \partial_{u_0}$,
its components $(0,2\int (u_{+}/u_0)dx)$ 
cannot be expressed in the form of a matrix times $(u_{+},u_0)$,
in contrast to the generators 
$\X=iu\partial_u$ in the case of timelike curves 
and $\X=ju\partial_u$ in the case of spacelike curves. 

The preceding properties can be summarized as follows. 

\begin{prop}
For null curves in $\Mink{2}$, 
with an arbitrary smooth parameterization, 
a minimal-isotropy null frame \eqref{nullparallelframe} is related 
to a Frenet null frame \eqref{Frenetnullframe}
by a little-group gauge transformation \eqref{gaugegroup}--\eqref{SonE}
which constitutes a Hasimoto transformation,
with the group parameter given by a Riccati equation \eqref{gaugerel1}. 
The Cartan matrix \eqref{nulltangentUWgauge} in a minimal-isotropy gauge 
belongs to a representation of the Lie algebra $\mk{so}(2,1)$ 
and has the algebraic characterization \eqref{gauge}
that it has no components in the little-group subalgebra that preserves the null tangent vector. 
Its non-vanishing components $(u_{+},u_0)$ 
are expressed in terms of the Frenet curvature and torsion $(\kappa,\tau)$ 
through a representation \eqref{curvtorssys} of the little group in $SO(2,1)$. 
Geometrically, 
these components $u_{+}$ and $u_0$ represent a little-group invariant and covariant of the null curve, relative to its arbitrarily fixed parameterization. 
\end{prop}

\section{Geometric flows of elastic null curves}
\label{sec:flows}

Each integrable system in the hierarchy of nonlinear Burgers-Airy systems 
shown in \Thmrefs{Burgersoddhierarchy}{Burgersevenhierarchy},
as well as in the hierarchies of semilinear wave systems and quasilinear Schr\"odinger systems show in \Thmref{additionalhierarchies}, 
determines an elastic null curve flow \eqref{flowvec}
through the relations \eqref{h+hperp}--\eqref{h-} and \eqref{Burgershierarchy}. 

We will now write these curve flows in terms of the Frenet null frame  \eqref{Frenetnullframe}--\eqref{Frenetnullframe-rel2}
by inverting the little-group gauge transformation given by 
equations \eqref{SonE}--\eqref{nulltangentframe} and equation \eqref{gaugerel1}. 

First, 
since both the tangent vector $\pos_x=\A^\t\wtil\E$ 
and the flow vector $\pos_t=\wtil\H^\t\wtil\E$ 
are gauge invariant, 
the inverse transformation acts on the components of the flow vector by 
\begin{equation}\label{flowtransformation}
\wtil\H \to \H = \S(\epsilon)^\t\wtil\H
\end{equation}
with 
\begin{equation}
\H = 
\bmatr
h^{+} \\
h^{-} \\
h^{\perp}
\ematr
\end{equation}
where 
\begin{equation}
\epsilon = -u_0/u_{+}
\end{equation}
from equation \eqref{u0u+}. 
The transformed components are given by 
\begin{equation}
h^{-}= h_{-}, 
\quad
h^{+} = h_{+} -(u_0/u_{+})h_{\perp} + \tfrac{1}{2}(u_0/u_{+})^2 h_{-},
\quad
h^{\perp} = h_{\perp} -(u_0/u_{+})h_{-} .
\end{equation}
Then the flow vector $\pos_t=\wtil\H^\t\wtil\E = \H^\t\E$ takes the form 
\begin{equation}
\begin{aligned}
\pos_t & = h^{+}\e_{+} +h^{-}\e_{-} +h^{\perp}\e_{\perp}
\\
& = (h_{+} -(u_0/u_{+})h_{\perp}+ \tfrac{1}{2}(u_0/u_{+})^2 h_{-}) \e_{+} 
+(h_{\perp} -(u_0/u_{+})h_{-}) \e_{\perp} 
+ h_{-} \e_{-} .
\end{aligned}
\end{equation}
Next, we use relations \eqref{h+hperp}--\eqref{h-}
to express $(h_{+},h_{-},h_{\perp})$ in terms of the flow components $(f_1,f_2)$
which are given by expression \eqref{Burgershierarchy} 
in terms of the variables $(u_{+},u_0)$. 

This leads to the following result. 

\begin{lem}\label{curveflowhierarchy}
Every two-component system 
\begin{equation}\label{f1f2}
\bmatr u_0 \\ u_{+} \ematr_t = \bmatr f_1 \\ f_2 \ematr 
\end{equation}
determines a null curve flow
\begin{equation}\label{nullcurveflow}
\pos_t = 
\big( D_x^{-1} f_1 -(u_0/u_{+}^2)(D_x-\tfrac{1}{2}u_0)F \big)\T
+\big( (1/u_{+})D_xF \big)\N -F\B
\end{equation}
with 
\begin{equation}
F= (D_x-u_0)^{-1}\big(u_{+}(D_x-u_0)^{-1}(f_2 -u_{+}D_x^{-1}f_1)\big)
\end{equation}
where $(\T,\N,\B)$ is a Frenet null frame \eqref{Frenetnullframe}--\eqref{Frenetnullframe-rel2}
relative to an arbitrary smooth parameterization $x$ of the null curve. 
The Frenet curvature $\kappa$ and the Frenet torsion $\tau$ of the null curve 
are determined from the variables $(u_0,u_+)$ 
by the inverse Hasimoto transformation \eqref{curvtorssys}. 
\end{lem}

When $f_1$ and $f_2$ are given by functions of the variables $(u_0,u_+)$ and their $x$-derivatives, 
the components of the resulting flow vector \eqref{nullcurveflow} 
for the parameterized null curve 
can be expressed in terms of $(\kappa,\tau)$ and their $x$-derivatives
through the Hasimoto transformation \eqref{u0u+}. 
In particular, 
this transformation determines $(u_0,u_+)$ 
up to a residual one-parameter gauge freedom \eqref{Ucovariance} 
which acts as a rigid representation of the little group in $SO(2,1)$, 
as discussed in \Secref{sec:gauge}. 
Consequently, the flow vector \eqref{nullcurveflow} will, in general, 
involve an arbitrary constant parameter coming from this residual freedom. 
This will determine a one-parameter family of flows for the null curve. 
These one-parameter flows will describe geometrical motions in the sense that 
their equation of motion \eqref{nullcurveflow} will be invariant 
under the isometry group $ISO(2,1)\simeq SO(2,1)\ltimes \Mink{2}$ of Minkowski space,
since $\kappa$ and $\tau$ are invariants of the parameterized null curve. 

Moreover, 
when a flow \eqref{nullcurveflow} arises from an integrable system, 
the resulting geometrical motion will also have the same integrability properties as the system. 

Any geometric null curve flow will induce a dynamical evolution 
on the Frenet curvature and torsion $(\kappa,\tau)$ of the parameterized null curve. 
To derive these evolution equations, 
we use the Hasimoto transformation \eqref{curvtorssys} relating $(\kappa,\tau)$ and $(u_+,u_0)$
to transform the components $(f_1,f_2)$ of the vector field 
$\X = f_1\partial_{u_0} + f_2\partial_{u_{+}} = F_1\partial_{\tau} + F_2\partial_{\kappa}$. 
This yields
\begin{equation}\label{curvtorssysXcomponents}
\bmatr F_1 \\ F_2 \ematr
= \Iop \bmatr f_1 \\ f_2 \ematr
\end{equation}
where 
\begin{equation}
\Iop = 
\bmatr
-(D_x + u_0)\kappa^{-1} & u_0\kappa^{-1}D_x\kappa^{-1} -\tau\kappa^{-1} \\
0 & 1
\ematr . 
\end{equation}
Hence, the evolution equations for $(\kappa,\tau)$ are given by 
\begin{equation}\label{curvtorsevoleqns}
\bmatr \tau \\ \kappa \ematr_t
= \bmatr F_1 \\ F_2 \ematr
= \Iop \bmatr f_1 \\ f_2 \ematr . 
\end{equation}

\subsection{Elastic flows on invariants}

Any null curve $\pos(x)$ in $\Mink{2}$ possess two geometrical invariants \cite{Bon}. 
The first invariant is the pseudo-arclength $ds = \eta(\T_x,\T_x)^{1/4}dx$,
where $\T=\pos_x$ is the null tangent vector. 
The second invariant is the null curvature $k = \eta(\what\N_s,\what\B)$, 
which is defined in terms of a geometric null Frenet frame $(\what\T,\what\N,\what\B)$,
with $\what\T = \pos_s$ being the pseudo-arclength parameterized null tangent vector;
$\what\N=\what\T_s$ being the principal normal unit vector;
and $\what\B$ being the null opposite vector relative to $\what\T$ and $\what\N$,
namely, $\eta(\what\B,\what\B)=0$ and $\what\T\times\what\B = \what\N$. 

These invariants $(s,k)$ are related to the Frenet curvature and torsion $(\kappa,\tau)$ of the null curve $\pos(x)$ 
by 
\begin{equation}
s=\txtint \sqrt{\kappa}\,dx,
\quad
k = \tau -\tfrac{1}{2}( (\kappa^{-1})_{xx} - ((\kappa^{-1/2})_x)^2 )
\end{equation}
which can be derived by looking for invariants $\chi(\tau,\kappa,\kappa_x,\kappa_{xx})$
under smooth reparameterizations $x\to \tilde x=f(x)$ with $f'(x)>0$, 
using the induced transformation \eqref{reparam} of $\tau$ and $\kappa$. 

For any null curve flow given by \Lemref{curveflowhierarchy}, 
the corresponding evolution equations \eqref{curvtorsevoleqns} on $(\kappa,\tau)$ 
induce evolution equations on the invariants $(s,k)$. 
Similarly to the derivation of equations \eqref{curvtorsevoleqns}, 
we obtain 
\begin{gather}
k_t = 
\tfrac{1}{2}(D_x+u_0)^2(\kappa^{-2}f_2) -u_0(\kappa^{-2}f_2)_x 
- \tfrac{1}{8}( (\kappa^{-2})_xf_{2x} + 3((\kappa^{-1})_x)^2 f_2 )
-(D_x+u_0)(\kappa^{-1}f_1) , 
\label{keqn}
\\
s_t =\tfrac{1}{2}\txtint \kappa^{-1/2} f_2\, dx . 
\label{seqn}
\end{gather}
Hence, the pseudo-arclength $s$ will not be preserved whenever 
a null curve flow \eqref{nullcurveflow} has $f_2\neq 0$.

\subsection{Geometrical Burgers flow and Airy flow}

We will now work out the explicit geometrical flows arising from 
the four main integrable systems derived in \Secref{sec:results}.
Recall, from \Thmref{structurethm}, 
we have $u_{+}=\kappa$ and $u_0 = -\epsilon(\kappa,\tau)\kappa$,
where $\epsilon(\kappa,\tau)$ is the general solution of the Riccati equation \eqref{gaugerel1}.

Writing the integrable Burgers system \eqref{Burgerssys} in the form \eqref{f1f2},
we obtain 
\begin{equation}
\begin{aligned}
f_1 & = u_0\kappa^{-1}\kappa_{xx} +(\tfrac{1}{2}u_0^2\kappa^{-1}-2\tau)\kappa_x
-\kappa \tau_x  -u_0\tau\kappa ,
\\
f_2 & = \kappa_{xx} +2u_0\kappa_x -2\tau\kappa^2 -u_0^2\kappa .
\end{aligned}
\end{equation}
Since $f_2$ is non-zero, 
this yields a family of elastic null curve flows
\begin{equation}\label{Burgerssyscurveflow}
\pos_t = -\tfrac{1}{2} u_0^2\kappa^{-2}\alpha \T + (\kappa + u_0\kappa^{-1}\alpha)\N -\alpha\B,
\quad
\alpha = (D_x -u_0)^{-1}(\kappa^2) . 
\end{equation}
The evolution equations induced on the Frenet curvature and torsion consist of 
\begin{align}
&\begin{aligned}
\kappa_t & = \kappa_{xx} + 2u_0 \kappa_x  -2\kappa^2\tau -\kappa u_0^2 , 
\end{aligned}
\\
&\begin{aligned}
\tau_t & = \tau_{xx} + (u_0^2 \kappa^{-2} +2\tau\kappa^{-1})\kappa_{xx} 
-2\tau\kappa^{-2} \kappa_x + 2\kappa_x\tau_x \kappa^{-1}
\\&\qquad
-2u_0(u_0^2\kappa^{-2} +\tau\kappa^{-1})\kappa_x 
+\tfrac{1}{4}\kappa^{-1}(2\kappa\tau+3u_0^2)(2\kappa\tau+u_0^2) . 
\end{aligned}
\end{align}

Next, writing the integrable nonlinear Airy system \eqref{Airysys} in the form \eqref{f1f2},
we obtain 
\begin{equation}
\begin{aligned}
f_1 & = u_0\kappa^{-1}\kappa_{xxx} +(u_0^2\kappa^{-1} -3\tau)\kappa_{xx} -\kappa\tau_{xx} +\tfrac{3}{2}u_0^2\kappa^{-2}\kappa_x^2 -3\tau_x\kappa_x 
-u_0(7\tau+\tfrac{3}{2}u_0^2\kappa^{-1})\kappa_x 
\\&\qquad 
-2u_0\kappa \tau_x  +2\tau^2\kappa^2 +u_0^2\tau\kappa , 
\\
f_2 & = \kappa_{xxx} +3u_0\kappa_{xx} +3u_0\kappa^{-1}\kappa_x^2 -3(u_0^2+3\tau\kappa)\kappa_x -3\kappa^2\tau_x . 
\end{aligned}
\end{equation}
Again, since $f_2$ is non-zero, 
this yields a family of elastic null curve flows, 
given by 
\begin{equation}\label{Airysyscurveflow}
\pos_t = -(\kappa\tau + \tfrac{3}{4}u_0^2\kappa^{-2})\T + (\kappa_x + \tfrac{3}{2}(u_0\kappa+\beta))\N -(\tfrac{1}{2}\kappa^2 + \tfrac{3}{2}\beta)\B,
\quad
\beta = (D_x -u_0)^{-1}(u_0\kappa^2) .
\end{equation}
The evolution equations induced on the Frenet curvature and torsion are given by 
\begin{align}
&\begin{aligned}
\kappa_t & = \kappa_{xxx}+3u_0\kappa_{xx}+3u_0\kappa_{x}^2\kappa^{-1} -3(3\tau\kappa +u_0^2)\kappa_x -3\kappa^2\tau_x , 
\end{aligned}
\\
&\begin{aligned}
\tau_t & = \tau_{xxx} +3(\tau\kappa^{-1}+\tfrac{1}{2}u_0^2\kappa^{-2})\kappa_{xxx}
+3\kappa^{-1}\kappa_x \tau_{xx}
+3\big( \kappa^{-2}(u_0^2\kappa^{-1}-\tau)\kappa_x +2\kappa \tau_x 
\\&\qquad
-u_0\kappa^{-1}(\tau + u_0^2\kappa^{-1}) \big)\kappa_{xx}
-\tfrac{3}{2}u_0^2\kappa^{-4}\kappa_x^3 
-3(\kappa^{-2}\tau_x +u_0\tau\kappa^{-2} +u_0^3\kappa^{-3}) \kappa_x^2
\\&\qquad
+6 u_0^2\kappa^{-1}(\tau +\tfrac{3}{8} u_0^2\kappa^{-1})\kappa_x -3\kappa\tau_x\tau .
\end{aligned}
\end{align}

Both of these elastic curve flows and corresponding evolution equations 
possess a hierarchy of symmetries given by the vector fields
\begin{equation}
\X^{(n)} = f_1^{(n)}\partial_{u_0} + f_2^{(n)}\partial_{u_{+}} = F_1^{(n)}\partial_{\tau} + F_2^{(n)}\partial_{\kappa}
\end{equation}
where 
\begin{equation}
\bmatr F_1^{(n)} \\ F_2^{(n)} \ematr
= \Iop \bmatr f_1^{(n)} \\ f_2^{(n)} \ematr 
= (\Iop\Rop\Iop^{-1})^n 
\bmatr \tau_x \\ \kappa_x \ematr,
\quad
n=0,1,2,\ldots 
\end{equation}
with $\X^{(0)}$ generating an $x$-translation symmetry.

\subsection{Additional geometrical flows}

In a similar way, 
the integrable semilinear wave system \eqref{wavesys1}--\eqref{wavesys2}
and the integrable quasilinear Schr\"odinger system \eqref{Schrsys}
yield the respective elastic null curve flows
\begin{equation}\label{wavesyscurveflow}
\begin{aligned}
\pos_t & = 
\big( e^{-v_0}\kappa + u_0\kappa^{-1}(e^{v_0}v_{+}^2)_x - e^{v_0}u_0^2\kappa^{-1} -\tfrac{1}{2} u_0^2\kappa^{-2}\zeta \big)\T 
\\&\qquad
+\big(e^{v_0}u_0 -v_{+}(e^{v_0}v_{+})_x  +u_0\kappa^{-1}\zeta\big)\N 
-\zeta\B
\end{aligned}
\end{equation}
with
\begin{equation}
\zeta = e^{v_0}D_x^{-1}\big( u_0\kappa +\tfrac{1}{2}(v_{+}^2)_x(e^{v_0}v_{+})_x \big)
\end{equation}
and
\begin{equation}\label{Schrsyscurveflow}
\begin{aligned}
\pos_t & = 
\big( \kappa^{-1}(e^{-v_0}v_{+})_x\big( (e^{v_0}\kappa)_x + e^{v_0}v_{+}(e^{v_0}u_0)_x \big) -\tfrac{1}{2} u_0^2\kappa^{-2}\nu \big) \T 
\\&\qquad
+\big( (v_{+}^2-1)(e^{v_0}u_0)_x  +e^{-v_0}v_{+} (e^{v_0}u_{+})_x+u_0\kappa^{-1}\nu \big)\N 
-\nu\B
\end{aligned}
\end{equation}
with
\begin{equation}
\nu = e^{v_0}D_x^{-1}\big( (v_{+}^2-1)\kappa(e^{v_0}u_0)_x +\tfrac{1}{2}(v_{+}^2)_x(e^{v_0}\kappa)_x \big)
\end{equation}
Here $v_0=-D_x^{-1}(\epsilon(\kappa,\tau)\kappa)$ 
and $v_{+}=D_x^{-1}(e^{-v_0}\kappa)$ are the potentials \eqref{u0potential} and \eqref{u+potential}
expressed in terms of $\kappa$ and $\tau$, 
where $\epsilon(\kappa,\tau)$ denotes the general solution of the Riccati equation \eqref{gaugerel1}.

\section{Conclusion}
\label{sec:conclude}

There are several interesting aspects and consequences of the main results in this paper. 

An explicit geometrical origin for Burger's systems, nonlinear Airy systems, 
and quasilinear NLS-type systems has been derived 
from elastic flows of null curves in Minkowski space. 
More importantly, 
the integrability structure of these two-component systems has been shown to arise 
through a Cole-Hopf transformation 
which is encoded in the Serret-Frenet equations of a new frame 
consisting of a null-vector counterpart of a parallel frame. 
This provides a new geometrical realization for Cole-Hopf transformations. 

Hence, the deep connection between nonlinear integrable systems and geometrical inelastic curve flows in Euclidean space 
extends to elastic null curve flows in Minkowski space.

Our methods and results can be developed in at least two interesting directions. 

First, 
elastic flows of null curves can be studied in $n$-dimensional Minkowski space 
$\Mink{n-1}$ for $n\geq 4$. 
The new analog of the Hasimoto transformation 
and the associated null frame which we have introduced in \Thmref{structurethm} 
can be generalized from $n=3$ dimensions to higher dimensions, 
similarly to the Euclidean case where 
the Hasimoto transformation and the associated parallel frame 
in three-dimensional Euclidean space 
have a natural generalization to all higher dimensions \cite{SanWan}. 
For $n\geq4$, 
the resulting null frame structure equations in $\Mink{n-1}$ will encode 
a hereditary recursion operator along with a Cole-Hopf transformation
which can be used to derive multi-component versions of the three main hierarchies of 
$C$-integrable nonlinear systems obtained in \Thmrefs{Burgersevenhierarchy}{Burgersoddhierarchy} and \Thmref{additionalhierarchies}. 
In addition, these integrable systems will correspond to elastic null curve flows
which possess a recursion operator and a hierarchy of symmetries. 

Second, 
a generalization from $n$-dimensional Minkowski space 
to curved Lorentzian symmetric spaces 
$M=G/SO(n-2,1)$, with $G=SO(n-1,1)$ and $G=SU(n-2,1)$ for $n\geq 3$ 
can be considered. 
Both spaces have the same $SO(n-1,1)$ gauge group for their frame bundles
as in the flat space $\Mink{n-1}$,
with the first space having constant-curvature while the other space has only constant scalar curvature. 
Elastic null curve flows in these curved spaces can be studied by the methods
we have used in this paper,
and similar interesting results can be expected to be obtained. 

A different direction of work than what we have considered in the present paper 
is the study of flows of differential invariants of null curves
in $\Mink{2}$ and $\Mink{3}$ \cite{MusNic,AndMar}
such that the curve is non-stretching with respect to its pseudo-arclength. 
Such flows also lead to integrable systems but which are of KdV type.

\appendix
\section{Minkowski space $\Mink{2}$}

A general introduction to the geometrical structure of Minkowski space 
can be found in \Ref{ONei}. 
The essential aspects of this structure needed for studying null curves 
will be summarized here. 

Three-dimensional Minkowski space is a vector space $\Mink{2}$ 
equipped with a Lorentz-signature $(-1,1,1)$ metric $\eta$
and a compatible volume form $\epsilon$. 

A vector $\vec v\in \Mink{2}$ is respectively 
{\em timelike}, {\em spacelike}, or {\em null} if 
its Minkowski norm $\eta(\vec v,\vec v)$ is negative, positive, or zero. 
The set of all null vectors comprises a $2$-dimensional surface 
through the origin in $\Mink{2}$, called the {\em lightcone}. 
A timelike vector $\vec v$ has unit norm if $\eta(\vec v,\vec v)=-1$,
while a spacelike vector $\vec v$ has unit norm if $\eta(\vec v,\vec v)=1$. 

The isometry group of Minkowski space $(\Mink{2},\eta)$ is the Poincar\'e group 
$ISO(2,1)\simeq SO(2,1)\ltimes \Mink{2}$,
which consists of time and space translations, rotations, and boosts. 
A subgroup of the Poincar\'e group that leaves invariant any fixed null vector
defines a {\em little group}. 

A vector cross-product $\times$ in Minkowski space is determined by the relation 
\begin{equation}
\eta(\vec v\times\vec w,\vec z)=\epsilon(\vec v,\vec w,\vec z)
\end{equation}
holding for all vectors $\vec v,\vec w,\vec z\in \Mink{2}$.
This cross-product obeys the basic properties
\begin{equation}\label{crossprod-prop1}
\vec v\times\vec w = - \vec w\times\vec v,
\quad
\eta(\vec v\times\vec w,\vec v) =0,
\quad
\eta(\vec v\times\vec w,\vec v\times\vec w) = \eta(\vec v,\vec w)^2 - \eta(\vec v,\vec v)\eta(\vec w,\vec w) 
\end{equation}
Another useful property is 
\begin{equation}\label{crossprod-prop2}
(\vec v\times\vec w)\times\vec z =\eta(\vec w,\vec z)\vec v -\eta(\vec v,\vec z)\vec w
\end{equation}
Compared to the vector cross-product in three-dimensional Euclidean space, 
here the first two properties are unchanged while the third and fourth properties each differ by a minus sign on the right hand side. 

An orthonormal basis for Minkowski space consists of a unit timelike vector $\e_0$
and two mutually orthogonal unit spacelike vectors $\e_1,\e_2$ 
that are orthogonal to $\e_0$:
$\eta(\e_0,\e_0)=-1$, $\eta(\e_0,\e_i)=0$, $\eta(\e_i,\e_j)=\delta_{ij}$
where $\delta_{ij}$ denotes the Kronecker symbol and $i,j=1,2$. 
A basis is {\em oriented} with respect to the volume form if 
$\epsilon(\e_1,\e_2,\e_0)=1$. 

The following two lemmas will be useful in the construction of moving frames 
for null curves. 

\begin{lem}\label{lem1:nullvec}
Any null vector $\vec v\in \Mink{2}$ can be decomposed into a multiple of 
the sum of a unit timelike vector $\vec w$ and a unit spacelike vector $\vec z$, 
by choosing any Minkowski plane $\Mink{1}\subset \Mink{2}$ that contains $\vec v$. 
Then the null vector $\vec v= c(\vec w+\vec z)$, $c=\const$, 
lies on one side of the lightcone in this plane, 
and reflecting $\vec z$ relative to $\vec w$ yields a null vector $c(\vec w-\vec z)$ 
lying on the opposite side of the lightcone in the same plane. 
\end{lem}

\begin{lem}\label{lem2:nullvec}
The perp space of a null vector $\vec v\in \Mink{2}$ is comprised by 
all vectors $\vec w$ satisfying $\eta(\vec w,\vec v)=0$. 
Every vector in the perp space is a linear combination 
$\vec w=c_\parallel\vec v + c_\perp\vec z$, 
$c_\parallel,c_\perp=\const$, where $\vec z$ is a spacelike vector orthogonal to $\vec v$.
The span of all spacelike vectors in the perp space is a spacelike line 
$\Rnum\subset\Mink{2}$ orthogonal to $\vec v$.
\end{lem}

The proof of these two lemmas is a straightforward computation, 
using a standard orthonormal basis for $\Mink{2}$, and will be omitted.

\end{document}